\documentclass[aps,preprint,preprintnumbers,nofootinbib]{revtex4}
\usepackage{epsfig}
\usepackage{color}
\usepackage{amsmath}

\def\CC{{\cal C}}

\def\CL{{\cal L}}
\def\CO{{\cal O}}

\def\CV{{\cal V}}
\def\CU{{\cal U}}
\def\CM{{\cal M}}

\def\MeV{\mathop{\rm MeV}\nolimits}
\def\GeV{\mathop{\rm GeV}\nolimits}

\def\bar{\overline}
\def\hat{\widehat}
\def\tilde{\widetilde}

\def\Tr{\textrm{Tr}}
\def\Str{\textrm{Str}}

\def\bk{$B_K\;$}

\def\etal{{\it et al.}}

\def\spose#1{\hbox to 0pt{#1\hss}}
\def\ltapprox{\mathrel{\spose{\lower 3pt\hbox{$\mathchar"218$}}
 \raise 2.0pt\hbox{$\mathchar"13C$}}}
\def\gtapprox{\mathrel{\spose{\lower 3pt\hbox{$\mathchar"218$}}
 \raise 2.0pt\hbox{$\mathchar"13E$}}}
\def\inapprox{\mathrel{\spose{\lower 3pt\hbox{$\mathchar"218$}}
 \raise 2.0pt\hbox{$\mathchar"232$}}}

\preprint{UW/PT 05-19}
\begin{document}

\title{\bk in Staggered Chiral Perturbation Theory}

\author{Ruth S. Van de Water}
\email{ruthv@u.washington.edu}
\affiliation{Physics Department, University of Washington, Seattle, WA 98195-1560}
\author{Stephen R. Sharpe}
\email{sharpe@phys.washington.edu}
\affiliation{Physics Department, University of Washington, Seattle, WA 98195-1560}

\date{\today}

\begin{abstract}
We calculate the kaon B-parameter, $B_K$, to next-to-leading order in staggered chiral perturbation theory.  We find expressions for partially quenched QCD with three sea quarks, quenched QCD, and full QCD with $m_u = m_d \neq m_s$.  We extend the usual power counting to include the effects of using perturbative (rather than non-perturbative) matching factors. Taste breaking enters through the $O(a^2)$ terms in the effective action, through $\CO(a^2)$ terms from the discretization of operators, and through the truncation of matching factors. These effects cause mixing with several additional operators, complicating the chiral and continuum extrapolations.  In addition to the staggered expressions, we present \bk at next-to-leading order in continuum PQ$\chi$PT for $N_f=3$ sea quarks with $m_u = m_d \neq m_s$.
\end{abstract}

\maketitle

\section{Introduction}
\label{sec:intro}

Experimental measurements of CP violation can be used to extract information about the CKM matrix.  In particular, the size of indirect CP violation in the neutral kaon system, $\epsilon_K$, combined with theoretical input, places an important constraint on the apex of the CKM unitarity triangle \cite{CKMfit, UTfit}.  Because $\epsilon_K$ is well-known experimentally \cite{Eidelman:2004wy}, the dominant source of error in this procedure is the uncertainty in the lattice determination of the nonperturbative constant $B_K$, which parameterizes the $K^0-\bar{K^0}$ matrix element.  Because new physics would likely produce unitarity violation in the CKM matrix and additional CP-violating phases, a precise determination of $B_K$ will help to constrain physics beyond the standard model.    

Promising calculations with partially quenched staggered fermions are in progress \cite{Gamiz:2004qx,W_Lee}.\footnote{Calculations of \bk with dynamical domain wall fermions are also underway \cite{Dawson:2004gu}.}  Because staggered fermions are computationally cheaper than other standard discretizations, they allow simulations with the lightest dynamical quark masses currently available.  Unfortunately, staggered fermions come with their own source of error -- taste symmetry breaking.  Each lattice fermion flavor comes in four tastes, which are degenerate in the continuum.  The nonzero lattice spacing, $a$, breaks the continuum taste symmetry at $\CO(a^2)$, and the resulting discretization errors are numerically significant at present lattice spacings \cite{MILCf}.  Thus one must use the chiral perturbation theory functional forms for \emph{staggered} fermions in order to correctly perform the combined continuum and chiral extrapolations incorporating taste violations \cite{BA1, BA2, Us}.   It is clear, then, that the potential improvement in precision of the lattice determination of \bk is limited without the availability of the appropriate staggered $\chi$PT expression.    

In this paper we calculate \bk to next-to-leading order (NLO) in an extended version of staggered chiral perturbation theory (S$\chi$PT) that incorporates perturbative operator mixing.  In the NLO S$\chi$PT expressions for quantities calculated thus far, e.g $f_K$ and $f_D$, taste-breaking primarily enters through additive corrections to the tree-level masses of pions inside loops \cite{BA1, BA2, Aubin:2004xd}.  The calculation of \bk in S$\chi$PT provides a more complicated example of taste-breaking in which new operators in the staggered chiral effective theory significantly change the result from that in continuum $\chi$PT, even at NLO.  The bulk of our new work is in correctly enumerating all possible operators which contribute.  It turns out that the NLO expression is in terms of 37 low-energy constants (21 for a single lattice spacing), 5 of which are already known from fitting other staggered lattice data.  Because fitting so many parameters requires a large amount of data, we show how to extract additional coefficients using simple kaon matrix elements.  Nevertheless, performing the continuum and chiral extrapolations of \bk will be challenging.     

\bigskip

We aim to keep the body of this paper accessible to those wishing to use our NLO S$\chi$PT expressions for calculations of \bk on the lattice.  Our paper is therefore organized as follows.  In Section~\ref{sec:BKStag} we define \bk in QCD, and discuss modifications to the quark-level operator necessary for its calculation with staggered quarks.  We summarize the main results of our \bk operator enumeration in Section~\ref{sec:SChPT}.  In Section~\ref{sec:BK444} we calculate \bk at 1-loop for a partially quenched (PQ) theory with three dynamical quark flavors (each with four tastes).  We then add the NLO analytic terms and give results for \bk to NLO in quenched, partially quenched, and full QCD in Section~\ref{sec:Res}.  Finally, in Section~\ref{sec:Lat} we explicitly show, for a single lattice spacing, how to determine six of the undetermined NLO coefficients, as well as how to use the S$\chi$PT fit results to extract the renormalization group invariant quantity $\hat{B}_K$ in the continuum. We conclude in Section~\ref{sec:Conc}.  As a corollary to our staggered result, we present the expression for \bk at NLO in \emph{continuum} PQ$\chi$PT with 2+1 sea quarks ($m_u=m_d\neq m_s$) in Appendix~\ref{app:contPQ}, since it is not available in the literature.  Appendix~\ref{app:SChPT} briefly reviews the essentials of staggered chiral perturbation theory necessary for understanding the details of our NLO calculation and operator enumeration.  Appendix~\ref{app:ops} enumerates in detail all of the operators which contribute to \bk at NLO.

\section{\bk with Staggered Quarks}
\label{sec:BKStag}

In this section we relate \bk in QCD to matrix elements of a continuum staggered theory in which one has taken the fourth-root of the quark determinant but there are still four tastes per valence quark flavor.  

\bigskip

The kaon $B$-parameter is defined as a ratio of matrix elements:
\begin{equation}
	\CM_K = \langle\bar{K^0} | \CO_K | K^0\rangle = B_K \CM_\textrm{vac} , 
\label{eq:BKdef}\end{equation}
where $\CO_K$ is a weak operator and $\CM_\textrm{vac}$ is the result for $\CM_K$ in the vacuum saturation approximation:
\begin{eqnarray}
	\CO_K &=& [\bar{s_a} \gamma_\mu (1 + \gamma_5) d_a][\bar{s_b}\gamma_\mu(1 + \gamma_5) d_b] \\
	\CM_\textrm{vac} &=& \frac{8}{3} \langle \bar{K^0} |[\bar{s_a} \gamma_\mu (1 + \gamma_5) d_a]|0\rangle \langle 0|[\bar{s_b}\gamma_\mu(1 + \gamma_5) d_b]| K^0\rangle . \label{eq:MVac} 
\end{eqnarray}
Note that there are separate summations over color indices $a$ and $b$.  Thus the matrix element $\CM_K$ receives contributions from two different quark-level contractions, one of which produces a single color loop and the other which has two color loops.  The expression in Eq.~(\ref{eq:BKdef}) can be simplified using the fact that $\CM_\textrm{vac}$ is related to the square of the kaon decay constant:
\begin{equation}
	\CM_K =\frac{8}{3} B_K m_K^2 f_K^2 \,,
\end{equation}
where we use the normalization that $f_K \approx 156 \MeV$.

In order to calculate \bk with \emph{staggered} fermions we must introduce the \emph{taste} degree of freedom, both in the operators ($\CO_K$) and in the states ($K^0$ and $\bar{K^0}$).  We choose the external staggered kaons to be taste $P$, by which we mean that the lattice meson operator contains the pseudoscalar taste matrix $\xi_5$.  Since this is the lattice Goldstone taste, its correlation functions satisfy $U(1)_A$ Ward identities, so S$\chi$PT expressions for its mass, decay constant, and other physical quantities are simpler than those for other tastes of PGBs.  In addition, it is a local kaon on the lattice, and therefore relatively simple to implement numerically.  Because the external kaons are taste $P$, the weak operator should also be taste $P$:    
\begin{equation}
	\CO_K^{naive} = \big[\bar{s_a} \big(\gamma_\mu (1 + \gamma_5) \otimes \xi_5\big) d_a\big]\big[\bar{s_b} \big(\gamma_\mu(1 + \gamma_5)\otimes \xi_5 \big) d_b\big]
\end{equation}  
However, when this operator is Fierz-transformed, not only do its color indices change, but it mixes with other tastes.  To remedy this, we introduce two types of valence quarks, 1 and 2, into the \bk operator:
\begin{eqnarray}
	\CO_K^{staggered} & = & 2 \bigg\{ \big[\bar{s1_a} \big(\gamma_\mu (1 + \gamma_5) \otimes \xi_5\big) d1_a\big]\big[\bar{s2_b} \big(\gamma_\mu(1 + \gamma_5)\otimes \xi_5 \big) d2_b\big] \nonumber \\
	& + & \big[\bar{s1_a} \big(\gamma_\mu (1 + \gamma_5) \otimes \xi_5\big) d1_b\big]\big[\bar{s2_b} \big(\gamma_\mu(1 + \gamma_5)\otimes \xi_5 \big) d2_a\big] \bigg\}
\label{eq:OKStag}\end{eqnarray} 
and take the matrix element between two types of kaons:
\begin{equation}
	\CM_K^{staggered} = \langle\bar{K^0_1}_P | \CO_K^{staggered} | {K^0_2}_P\rangle\,,
\label{eq:MKStag}\end{equation}
where $K^0_1$ is an $\bar{s1}d1$ meson and $K^0_2$ is an $\bar{s2}d2$ meson.  The extra valence quarks require explicit inclusion of both color contractions in $\CO_K^{staggered}$, while the overall factor of 2 ensures that $\CM_K$ and $\CM_K^{staggered}$ have the same total number of contractions.  When $\CO_K^{staggered}$ is Fierz-transformed, it has a new \emph{flavor} structure so it does not contribute to $\CM_K^{staggered}$.  Thus $\CM_K^{staggered}$ cannot receive contributions from incorrect tastes.  We choose that the two sets of valence quarks have equal masses, i.e. $m_{d1} = m_{d2} \equiv m_x$ and $m_{s1} = m_{s2} \equiv m_y$, in order to simplify future expressions.   

The continuum staggered theory differs from QCD in that it has four copies of each valence quark, so $\CM_K$ and $\CM_K^{staggered}$ are not precisely equal.  Nevertheless they are simply related by overall normalization factors.  Specifically, 
\begin{eqnarray}
	\CM_K & = & \frac{1}{N_{t}}\CM_K^{staggered} \\
	\langle 0| \bar{s} \gamma_\mu \gamma_5 d |K^0\rangle & = & \sqrt{\frac{1}{N_{t}}} \langle 0| \bar{s} (\gamma_\mu \gamma_5 \otimes \xi_5) d |K^0_P\rangle ,
\end{eqnarray}
where $N_t = 4$ is the number of tastes per valence quark flavor \cite{Kilcup:1997ye}.  We emphasize that these expressions assume that one has already taken the fourth-root of the quark determinant and the continuum limit, and that the $\sqrt[4]{\mbox{Det}}$ procedure is valid.  We now re-express \bk in terms of the staggered matrix elements calculated on the lattice:
\begin{equation}
  B_K = \frac{\frac{1}{N_{t}}\langle\bar{K^0_1}_P | \CO_K^{staggered} | {K^0_2}_P\rangle}{\frac{8}{3}\sqrt{\frac{1}{N_{t}}} \langle \bar{K^0}_P | \bar{s} (\gamma_\mu \gamma_5 \otimes \xi_5) d |0\rangle \sqrt{\frac{1}{N_{t}}} \langle 0| \bar{s} (\gamma_\mu \gamma_5 \otimes \xi_5) d |K^0_P\rangle}.
\label{eq:stagBKdef}\end{equation}
Because the various factors of $N_t$ cancel, \bk in QCD is \emph{equal} to the value of the analogous ratio of staggered matrix elements.

Current calculations of \bk with staggered quarks use the partially quenched approximation.  The quark content the corresponding PQ theory is quite large.  We have already seen that staggering requires two sets of $d$ and $s$ valence quarks.  Partial quenching adds two corresponding sets of ghost quarks as well as three sea quarks, resulting in \emph{eleven} total quark flavors, each of which comes in four tastes.  In order to make this completely clear, we show the explicit form of the quark mass matrix:
\begin{equation}
	M = diag\{\underbrace{m_uI,\, m_dI,\, m_sI}_\textrm{sea},\, \underbrace{m_{x}I,\, m_{y}I}_\textrm{valence 1},\, \underbrace{m_{x}I,\, m_{y}I}_\textrm{valence 2},\, \underbrace{m_{x}I,\, m_{y}I}_\textrm{ghost 1},\, \underbrace{m_{x}I,\, m_{y}I}_\textrm{ghost 2}\} ,
\end{equation}
where $I$ is the $4 \times 4$ identity matrix.  Thus the generalization of \bk to PQ staggered quarks is relatively straightforward conceptually, although nontrivial to implement on the lattice.  

\section{Generalized Staggered Chiral Perturbation Theory for \bk}
\label{sec:SChPT}

The previous section described how to calculate \bk with staggered quarks in the continuum, but our expression for \bk in S$\chi$PT must describe \bk at $a\neq0$ if it is to be used for continuum and chiral extrapolations of lattice data.  It is therefore necessary to discuss in some detail how \bk is actually calculated on the lattice.  

The \bk matrix element, $\CM_K^{staggered}$, receives a contribution from the lattice version of $\CO_K^{staggered}$, as well as from other lattice operators that are in the same representation of the symmetry group that maps a hypercube onto itself \cite{Verstegen}:  
\begin{eqnarray}
	\CO_K^{staggered, cont} &=& \CO_K^{staggered, lat} + \frac{\alpha}{4\pi}[\mbox{taste }P\mbox{ ops.}] + \frac{\alpha}{4\pi}[\mbox{other taste ops.}]\nonumber\\
	&& + \alpha^2[\mbox{all taste ops.}] + a^2[\mbox{all taste ops.}]  + \ldots ,
\label{eq:Mixing}\end{eqnarray}
where $\alpha$ is the strong coupling constant. The 1-loop perturbative matching coefficients between $\CO_K^{staggered}$ in the continuum and four-fermion lattice operators are known, and are numbers of order unity times $\alpha/4\pi$ \cite{LSMatch}.  However, the 2-loop matching coefficients have not been determined, so, in order to remain conservative, we consider them to be of order unity times $\alpha^2$ \emph{without} any factors of $4\pi$.  Current numerical staggered calculations are in fact of the following matrix element:
\begin{equation}
  \langle\bar{K^0_1}_P| \CO_{1-loop}(\mbox{taste }P) |{K^0_2}_P\rangle \equiv \CM_\textrm{lat},
\label{eq:LatOp}\end{equation}
where the subscript ``$1-loop$'' and the argument ``taste $P$'' indicate that one includes all staggered lattice operators with taste $P$ that mix with the latticized \bk operator at $\CO(\alpha/4\pi)$, i.e. those in the second term on the RHS of Eq.~(\ref{eq:Mixing}), using the appropriate matching coefficients.  However, one neglects wrong-taste and higher-order perturbative mixing (terms three and four), as well as all operators which arise through discretization effects (term five), in Eq.~(\ref{eq:Mixing}).  Although the expression in Eq.~(\ref{eq:LatOp}) differs from the continuum matrix element, it reduces to the desired quantity in the continuum limit.  Generically,
\begin{equation}
  \CM_\textrm{lat} = \CM_\textrm{cont} + \frac{\alpha}{4\pi}\CM' + \alpha^2 \CM'' + a^2 \CM''' + \ldots ,
\end{equation}
where $\CM_\textrm{cont}$ is the desired continuum result.  The matrix element $\CM'$ comes from neglecting taste-violating 1-loop operator mixing, while $\CM''$ comes from neglecting 2-loop operator mixing.  Both taste-breaking and taste-conserving discretization errors generate $\CM'''$.    

We are now in a position to determine the appropriate power-counting scheme for calculating \bk at next-to-leading order in S$\chi$PT.  Clearly it must incorporate $a^2$, $\alpha/4\pi$, and $\alpha^2$.  Continuum $\chi$PT is a low-energy expansion in both the pseudo-Goldstone boson (PGB) momentum and the quark masses;  it assumes that $p_{PGB}^2/\Lambda_{\chi}^2 \sim m_{PGB}^2/\Lambda_{\chi}^2$, where $m_{PGB}^2 \propto m_q$ and $\Lambda_\chi \approx 4\pi f_{\pi}$ is the $\chi$PT scale.  However, the numerical values of $m_q$, $a^2$ and $\alpha$ all depend on the particular parameters of a given lattice simulation.  Current PQ staggered lattice simulations \cite{MILCf} use a range of PGB masses from $m_{PGB}^2/\Lambda_\chi^2 = 0.04 - 0.2$, so our S$\chi$PT expression must apply throughout this range.  Generic discretization errors are of the size $a^2 \Lambda_\textrm{QCD}^2$, which is approximately $0.04$ for $1/a \sim 2\GeV$ and $\Lambda_\textrm{QCD} \sim 400 \MeV$, so they are comparable to the minimal $m_{PGB}^2/\Lambda_{\chi}^2$ and should be included at the same order.\footnote{Our generic discretization errors are not suppressed by $\alpha$ because we consider unimproved operators. This is in contrast to the mass and decay constant calculations of Ref.~\cite{MILCf}, which use improved operators and therefore have smaller generic discretization errors of around 2\%.}  Taste-breaking discretization errors, on the other hand, are caused by exchange of gluons with momentum $\pi/a$, and therefore receive an additional factor of $\alpha^2_V(\pi/a)$, so their size must be considered separately.\footnote{Taste violations are suppressed by $\alpha^2$ rather than $\alpha$ because we assume an improved action.}  At the lightest quark masses, the lattice Goldstone pion mass is comparable to the mass splittings among the other PGB tastes: $m_{PGB}^2/\Lambda_{\chi}^2 \sim a^2 \alpha_V^2(q^* = \pi/a) \Lambda^2 \sim 0.04$, where $\Lambda$ is a QCD scale (distinct from $\Lambda_\textrm{QCD}$) which turns out to be around $1200 \MeV$.  Thus they are not suppressed relative to pure $\CO(a^2)$ discretization effects by the additional powers of $\alpha$, as naive power-counting would suggest, because of the large scale $\Lambda$ associated with the taste-breaking process at the quark level \cite{Mason:2002mm}.  Standard S$\chi$PT only includes discretization effects -- we now consider additional errors from perturbative operator matching, which depend upon $\alpha_V(q^*)$.  Generically, the choice of $q^*$ is process-dependent, and the value of $\alpha_V(q^*)$ ranges from $\sim 0.3 - 0.55$ for $q^* = \pi/a - 1/a$ at $a=.125$fm \cite{CB_alpha}.  Thus both $\alpha/4\pi$ and $\alpha^2$ must be included at lowest-order in our power counting.\footnote{The quoted values of $\alpha$ indicate that it is perhaps necessary to include $\CO(\alpha^3)$ errors as well;  we assume that this is not the case.}

In light of this discussion, we adopt the following extended S$\chi$PT power-counting scheme:      
\begin{equation}
	p^2 \sim m \sim a^2 \sim a_\alpha^2 \sim \alpha/4\pi \sim \alpha^2\,,
\end{equation}
where $a_\alpha^2 \equiv \alpha^2_V(\pi/a) a^2$.  We account for the fact that $\alpha^2 a^2$ terms in the action are enhanced numerically by including them at the same order as simple discretization effects.\footnote{We assume that taste-conserving discretization errors are not enhanced since such behavior has not been observed in staggered lattice simulations.}  In fact, while it may seem ad hoc, $p^2 \sim m \sim a_\alpha^2$ is the \emph{standard} S$\chi$PT power counting scheme.  It is simply not traditionally written as such because standard S$\chi$PT calculations have only included $\CO(a_\alpha^2)$ taste-breaking discretization errors from the action, and have therefore not needed to contrast them with pure $\CO(a^2)$ discretization effects.  We also use conservative power-counting for the perturbative errors by assuming that 2-loop contributions are not significantly smaller than those from 1-loop diagrams.  We emphasize that our scheme is \emph{phenomenologically based} on the particular parameter values of current staggered simulations -- simulations using significantly lighter quark masses or smaller lattice spacings would require a different scheme and result in a different S$\chi$PT expression for \bk.   

We must determine all of the contributions at NLO in our power counting to $\CM_\textrm{lat}$, including the effects of perturbative matching and discretization errors.  We then find \bk by using $\CM_\textrm{lat}$ in the ratio given in Eq.~(\ref{eq:stagBKdef}).  In S$\chi$PT, $\CM_\textrm{lat}$ is simply the matrix element of a sum of operators with undetermined coefficients:
\begin{equation}
  \CM_\textrm{lat} = \langle \bar{K^0_1}_P | \CC_\chi^i \CO_\chi^i | {K^0_2}_P \rangle, 
\end{equation}
so our goal is really to determine all \emph{operators} in the chiral effective theory, $\CO_\chi^i$, that contribute at NLO to the above matrix element.  Once we do so, the full NLO S$\chi$PT expression will contain the LO tree-level term, 1-loop corrections, and analytic terms of the same order as the 1-loop terms.\footnote{Note the distinction between \bk at NLO and \bk at 1-loop, which does not contain the analytic terms.}  This task is not as daunting as it first seems given the extended power-counting scheme because it turns out that the only LO contribution to \bk is of $\CO(p^2)$.  Thus operators of the following orders in our power-counting contribute to \bk at NLO:
\begin{eqnarray}
	\textrm{1-loop:} && \CO(p^2), \;\CO(m), \;\CO(a^2), \;\CO(a_\alpha^2), \;\CO(\alpha/4\pi), \;\CO(\alpha^2) \nonumber\\
	\textrm{NLO analytic:} && \CO(p^4), \; \CO(p^2 m), \;\CO(p^2 a^2), \;\CO(p^2 a_\alpha^2), \;\CO(p^2 \alpha/4\pi), \;\CO(p^2 \alpha^2), \nonumber\\
	&& \CO(m^2), \;\CO(m a^2), \;\CO(m a_\alpha^2), \;\CO(m \alpha/4\pi), \;\CO(m \alpha^2),  \nonumber\\
	&& \CO(a^4), \;\CO(a_\alpha^4), \;\CO(\alpha^2/16\pi^2), \; \CO(\alpha^4) \,.
\end{eqnarray}

\bigskip

In the rest of this section we give an overview of the operators that contribute to \bk at 1-loop in S$\chi$PT.  Appendix~\ref{app:ops} contains the detailed operator enumeration as well as the determination of analytic terms. 

Clearly the lattice version of the continuum quark level operator $\CO_K^{staggered}$, Eq.~(\ref{eq:OKStag}), will generate the dominant contribution to \bk, so we map $\CO_K^{staggered}$ onto chiral operators first.  This is relatively straightforward;  we follow the graded group-theory method of Refs.~\cite{Me, Us} to insure that we include all possible linearly-independent operators.  This results in two operators, Eqs.~(\ref{eq:OK}) and (\ref{eq:ON}), the first of which is simply the staggered, partially quenched analog of the continuum $(27_L, 1_R)$ chiral operator, and the second of which arises because we use two sets of valence quarks.  Because both chiral operators come from the same quark-level operator, their coefficients are related, and turn out to be equal.  This is crucial because \bk in continuum $\chi$PT only depends on a single parameter \cite{BSW}, and our result must match the continuum one when $a\rightarrow0$.  Only the ``standard'' chiral operator contributes to \bk at tree-level, and its contribution is of $\CO(p^2)$.

Next we determine the chiral operators which contribute to \bk because of perturbative operator mixing.  Four-fermion operators which mix with $\CO_K^{staggered}$ but are \emph{not taste P} are unaccounted for in Eq.~(\ref{eq:LatOp}), so they introduce errors into \bk that are of $\CO(\alpha/4\pi)$, as well as errors of $\CO(\alpha^2)$.\footnote{Recall that $\alpha/4\pi \sim \alpha^2$ in our power-counting because we make no assumption about the numerical size of the $\CO(\alpha^2)$ matching coefficients.}  We therefore map them onto chiral operators with two undetermined coefficients of $\CO(\alpha/4\pi)$ and $\CO(\alpha^2)$, respectively.  The resulting eight chiral operators are given in Eqs.~(\ref{eq:O1I})~-~(\ref{eq:O3A}).  Taste $P$ four-fermion operators are accounted for at $\CO(\alpha/4\pi)$ using the matching coefficients, but they still introduce errors into \bk at $\CO(\alpha^2)$, through a new $\CO(\alpha^2)$ operator given in Eq.~(\ref{eq:O1P}). 

To match the continuum regularized operator, $\CO_K^{staggered,cont}$, we also need to include operators of higher dimension. From now on we will generically refer to this matching of continuum onto lattice operators as ``mixing".  In particular, we must include all dimension 8 operators that are in the same representation of the symmetry group that maps a hypercube onto itself.  Because dimension 8 operators are explicitly suppressed by a factor of $a^2$ relative to dimension 6 operators, they can map onto chiral operators with coefficients of $\CO(a^2)$ that contribute to \bk at NLO.  There turn out to be nine such operators -- the same as those which arise from perturbative operator mixing.

Finally, four-fermion operators from the staggered action that arise due to discretization effects can enter time-ordered products with the \bk operator $\CO_K^{staggered}$, effectively correcting the \bk four-fermion vertex at $\CO(a_\alpha^2)$.  The diagram for this process is shown in Figure~\ref{fig:ActIns}.  We must therefore determine the chiral operators which result from a \emph{combination} of the \bk four-fermion operator and the inserted four-fermion operator from the action.  We do so using the spurion method of Ref.~\cite{Us}.  This results in four new chiral operators, Eqs.~(\ref{eq:O2P})~-~(\ref{eq:O3T}), as well additional copies of seven operators which had appeared previously from operator mixing.

In total, we find fifteen chiral operators; we will determine their contributions to \bk at 1-loop in the following section.  We note, however, that many of these chiral operators arise in more than one way and correspond to more than one quark level operator, so they actually have \emph{more than one undetermined coefficient}.  We list the $\CO(\alpha/4\pi)$, $\CO(\alpha^2)$, $\CO(a^2)$, and $\CO(a_\alpha^2)$ operators and their coefficients in Table~\ref{tab:NLOCoeff}.  For example, the operator $\CO_\chi^{1A}$, which comes both from 1-loop perturbative matching and from insertions of operators in the staggered action has three coefficients, one multiplied by $\alpha/4\pi$, one multiplied by $\alpha^2$, and one multiplied by $a^2$.  The three coefficients can be separated in principle by carrying out simulations at more than one lattice spacing.  We emphasize that they \emph{cannot} be lumped together in a fit to multiple lattice spacings.  In following sections, including the final expressions for \bk at NLO, we only associate a single coefficient, $\CC^i_\chi$, which each operator.  This is simply to reduce the size of the final expressions.  One must combine the expressions with Table~\ref{tab:NLOCoeff} in order to make them complete.

\section{\bk at 1-Loop for 4+4+4 Dynamical Flavors}
\label{sec:BK444}

\begin{figure}
\begin{tabular}{ccccc}
	\epsfxsize=1.4in \epsffile{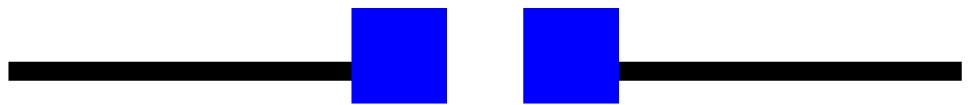} & $\;\;\;\;\;\;\;\;$ & \epsfxsize=1.4in \epsffile{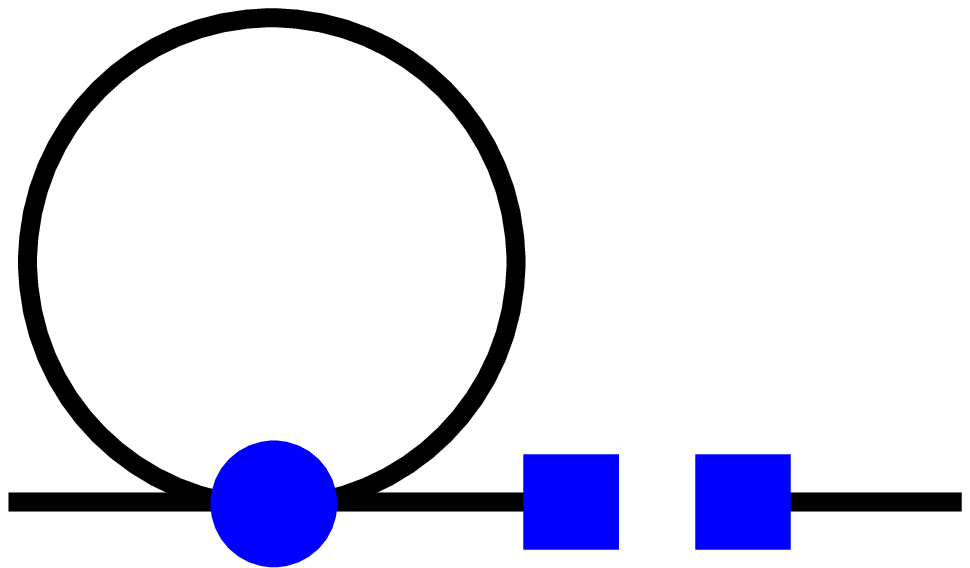} & $\;\;\;\;\;\;\;\;$ & \epsfxsize=1.4in \epsffile{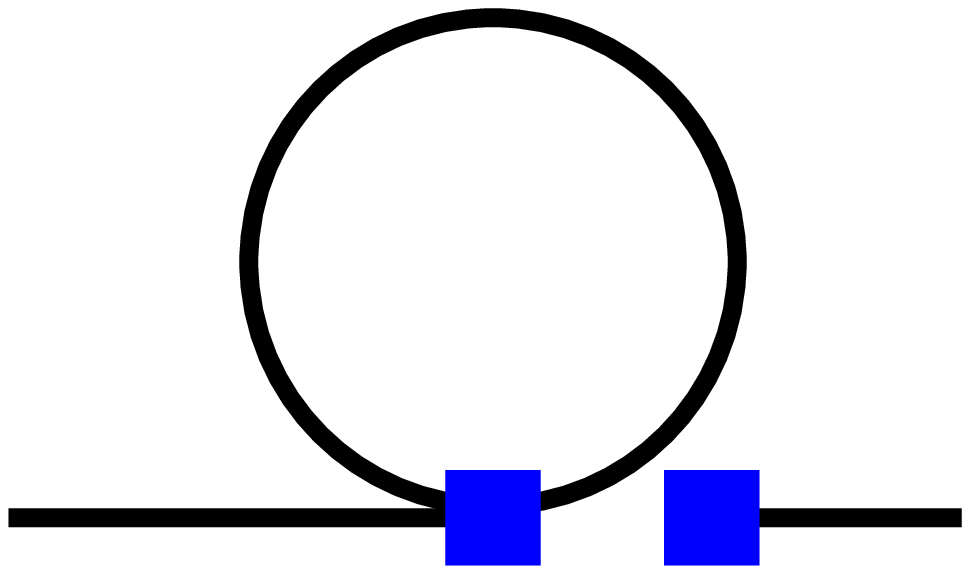} \\
	(a) & $\;\;\;\;\;\;$ & (b) & $\;\;\;\;\;\;$ & (c) \\
	\epsfxsize=1.4in \epsffile{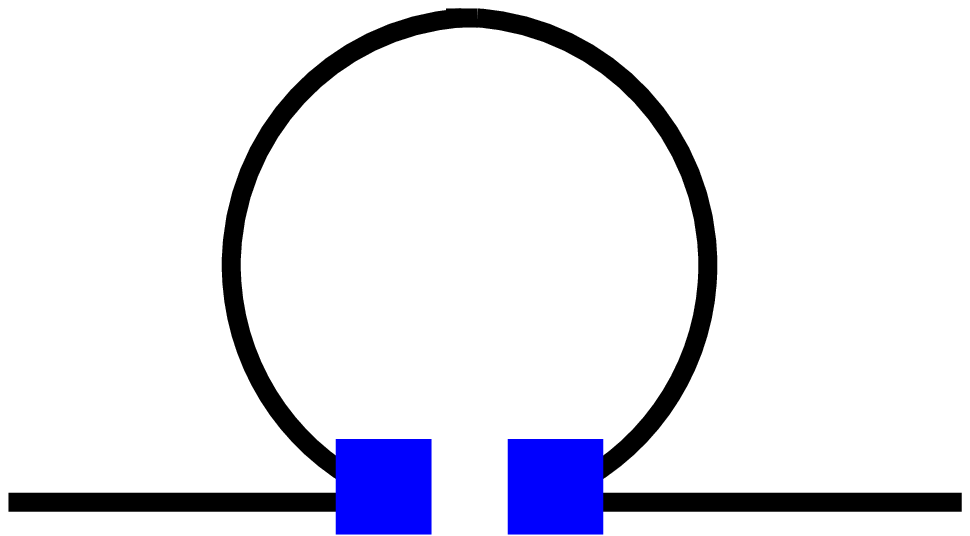} & $\;\;\;\;\;\;$ & \epsfxsize=1.4in \epsffile{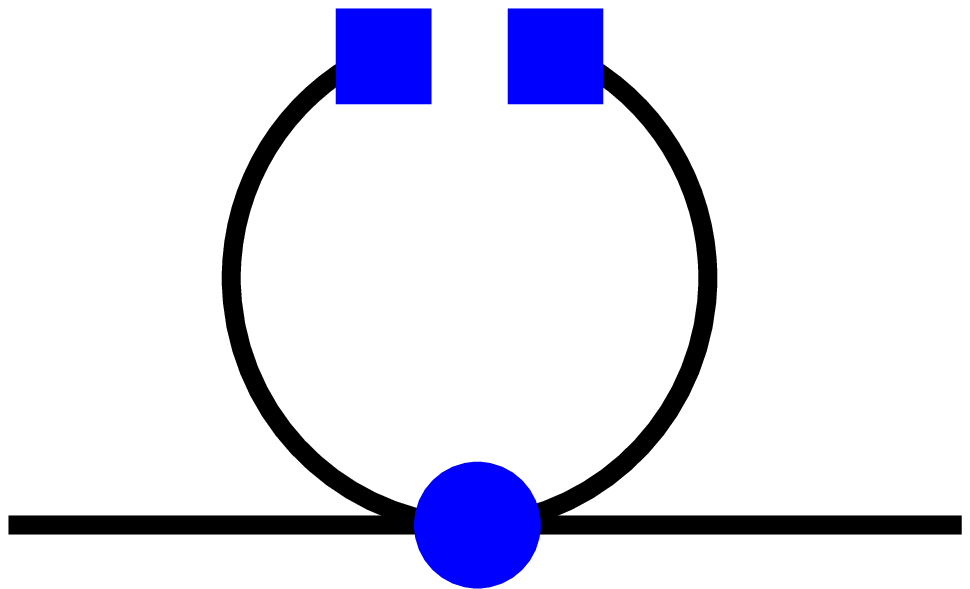} & $\;\;\;\;\;\;\;\;$ & \epsfxsize=1.4in \epsffile{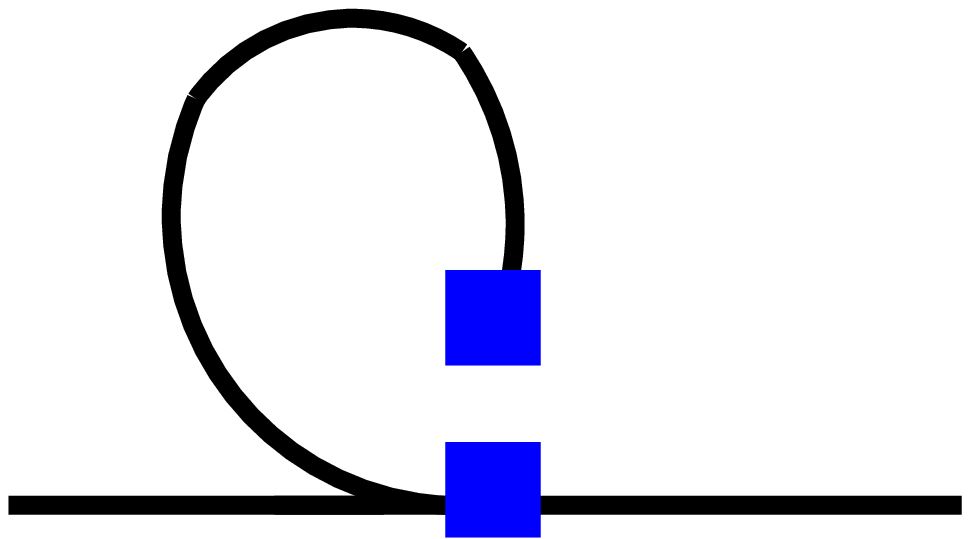} \\
	(d) & $\;\;\;\;\;\;$ & (e) & $\;\;\;\;\;\;$ & (f) \\
	\end{tabular}
	\caption{Tree-level and 1-loop contributions to $\CM_K$.  In diagrams (a) -- (e), one external PGB is a $\bar{K^0_1}$ and the other is a $K^0_2$.  Diagram (b) generates kaon wavefunction renormalization.  The circle represents a vertex from the LO staggered chiral Lagrangian.  The squares represent an insertion of the $B_K$ operator.  Each box ``changes" the quark flavor from $d \leftrightarrow s$.  Diagram (f) does not actually contribute in the staggered case because it has the wrong flavor structure to contract with two different external kaons.}\label{fig:BK1Loop}\end{figure}

We first calculate \bk at tree-level, which receives a contribution from a single diagram, Figure~\ref{fig:BK1Loop}(a), and from a single chiral operator:
\begin{equation}
	\CO_{\chi}^{K} = \sum_\mu \Big( \Str(\Sigma \partial_\mu \Sigma^\dagger F_1)\Str(\Sigma \partial_\mu \Sigma^\dagger F_2) + p.c. \Big).
\end{equation}
In the diagram, the \bk chiral operator is shown as two disconnected squares, each of which produces a kaon.  This reflects the structure of $\CO_{\chi}^{K}$, which has two supertraces in which the matrices $F_1$ and $F_2$ have the appropriate taste ($\xi_5$) and flavor ($s1\bar{d1}$ and $s2\bar{d2}$) structures to produce a $K^0_1$ and a $K^0_2$.  $\CO_{\chi}^{K}$ is just the staggered analog of the continuum $(27_L, 1_R)$ \bk operator -- it reduces to the standard form after removing the taste structure ($\xi_5 \rightarrow \xi_I$) and the partial quenching ($d1, d2 \rightarrow d$ and $\Str \rightarrow \Tr$).  The tree-level \bk matrix element is simple:
\begin{eqnarray}
	\CM_K^{LO} & = & - \frac{2 \CC_{\chi}^{K}}{f^2} \sum_\mu \langle\bar{K^0_1} \,|\,  \Str( \partial_\mu \pi F_1)\Str( \partial_\mu \pi F_2) \,|\, K^0_2\rangle \nonumber \\
	& = & - \frac{32 \CC_{\chi}^{K}}{f^2} \; \partial_\mu \bar{K^0_1} \; \partial_\mu K^0_2 \nonumber \\
	& = & \frac{32 \CC_{\chi}^{K}}{f^2} \; m_{xy_P}^2 , \nonumber \\
\end{eqnarray}
where $m_{xy_P}^2 = \mu(m_x + m_y)$ is the tree-level kaon mass.  The kaon B-parameter itself is just the ratio $\CM_K/\CM_\textrm{vac}$, where $\CM_\textrm{vac}$ is given in Eq.~(\ref{eq:MVac}).\footnote{Because we work only in the staggered theory for the rest of the paper, we drop the superscript ``staggered" from both operators and matrix elements in order to reduce clutter in expressions.}  Because the matrix element $\CM_\textrm{vac}$ is proportional to the square of the axial current, it is proportional to $f^2 p^2$ at tree-level:
\begin{equation}
	\CM_\textrm{vac}^{LO} = \frac{8}{3} f^2 m^2_{xy_P} . 
\end{equation}
Therefore the kaon mass drops out of the ratio of matrix elements:
\begin{equation}
	\left( \frac{\CM_K}{\CM_\textrm{vac}} \right)^{LO} = \; \frac{12}{f^4} \CC_\chi^K \; \equiv B_0 ,
\label{eq:B0}\end{equation}
and \bk is just a constant at lowest order in S$\chi$PT.  

\bigskip

We are now ready to calculate \bk at 1-loop in S$\chi$PT.  It is useful to simplify the form of \bk as much as possible \emph{diagrammatically} before explicitly showing any expressions.  In general, $\CM_K$ receives 1-loop contributions from the five diagrams shown in Figures~\ref{fig:BK1Loop}(b)--(e) \cite{Sharpe:1992ft}.    However, one can easily show that Figure~\ref{fig:BK1Loop}(f) cannot contribute in the staggered case because it is impossible to draw a corresponding quark-line diagram with two different external kaons.  Moreover, because \bk is defined by the \emph{ratio} $\CM_K/\CM_\textrm{vac}$, cancelations occur between matrix elements in the same manner as the kaon mass dropped out at tree-level.  It turns out that diagrams \ref{fig:BK1Loop}(b) and \ref{fig:BK1Loop}(c) cancel entirely in the \bk ratio, as we now show.  

The 1-loop matrix element $\CM_K$ can be expressed as
\begin{equation}
	\CM_K = \frac{8}{3} B_0 f^2 m_{xy_P}^2 \big\{1 + X[\mbox{Figs. \ref{fig:BK1Loop}(b)--(c)}]\big\} + X'[\mbox{Figs. \ref{fig:BK1Loop}(d)--(e)}]\,,
\end{equation}
where $X$ and $X'$ denote the results of the specified diagrams and $m_{xy_P}^2$ is the 1-loop kaon mass-squared.\footnote{Note that $X'$ is not within the curly braces because it comes from operators other than $\CO_\chi^K$ and is therefore not proportional to $B_0$.}  The NLO expression for $\CM_{vac}$ has the same form as the tree-level one:
\begin{equation}
	\CM_\textrm{vac} = \frac{8}{3} m_{xy_P}^2 f_{xy_P}^2\,,
\end{equation}
except that both $m_{xy_P}^2$ and $f_{xy_P}$ become the 1-loop quantities.  Now consider diagrams \ref{fig:BK1Loop}(b) and \ref{fig:BK1Loop}(c).  Visually, it is easy to see that they \emph{factorize} -- if one draws a vertical line between the boxes, the left halves of the two diagrams renormalize $f_K$ at 1-loop while the right halves just produce $f$ at tree-level.  It therefore seems natural that $X[\mbox{Figs. \ref{fig:BK1Loop}(b)-- (c)}]$ should be related to the 1-loop renormalization of the kaon decay constant, and, if one works out the details, one finds that 
\begin{equation}
	X[\mbox{Figs. \ref{fig:BK1Loop}(b)-- (c)}] = 2 \frac{\delta f_{NLO}}{f}\,,
\end{equation}
where the factor of two results from the fact that loop can be on either leg.  Thus diagrams (b) and (c) are exactly those necessary to turn the tree-level $f^2$ into the 1-loop $f_{xy_P}^2$:\footnote{Note that the NLO analytic contributions to $f_{xy_P}$ simply change the coefficients of the NLO analytic corrections to $B_K$.}
\begin{equation}
	\CM_K = \frac{8}{3} B_0 f_{xy_P}^2 m_{xy_P}^2  + X'[\mbox{Figs. \ref{fig:BK1Loop}(d)--(e)}]\,,
\end{equation}
and \bk only depends on diagrams \ref{fig:BK1Loop}(d) and \ref{fig:BK1Loop}(e):
\begin{equation}
	B_K^{1-loop} = B_0 + \frac{3}{8}\frac{X'[\mbox{Figs. \ref{fig:BK1Loop}(d)--1(e)}]}{f_{xy_P}^2 m_{xy_P}^2}\,.
\end{equation}

The diagrams in Figs.~\ref{fig:BK1Loop}(d) and \ref{fig:BK1Loop}(e) receive contributions from all of the enumerated \bk chiral operators, which we list here for convenience:
\begin{eqnarray}
	\CO_{\chi}^{K} & = &  \sum_\mu \Big( \Str(\Sigma \partial_\mu \Sigma^\dagger F_1)\Str(\Sigma \partial_\mu \Sigma^\dagger F_2) + \Str(\Sigma^\dagger \partial_\mu \Sigma F_1)\Str(\Sigma^\dagger \partial_\mu \Sigma F_2) \Big) \nonumber\\
	\CO_{\chi}^{N} & = &  \sum_\mu \Big( \Str(\Sigma \partial_\mu \Sigma^\dagger F_1 \Sigma \partial_\mu\Sigma^\dagger F_2) + \Str(\Sigma^\dagger \partial_\mu \Sigma F_1 \Sigma^\dagger \partial_\mu \Sigma F_2) \Big) \nonumber\\
	\CO_{\chi}^{1P} & = &  \Str(F_1 \Sigma F_2 \Sigma^\dagger) \nonumber\\[2mm]
	\CO_{\chi}^{2P} &=&  \Str(\xi_5  \Sigma F_1 \Sigma^\dagger \xi_5  \Sigma F_2 \Sigma^\dagger) + p.c. \nonumber\\[2mm]
	\CO_{\chi}^{3P} &=&  \Str(\xi_5 \Sigma F_1 \Sigma^\dagger)\Str(\xi_5 \Sigma F_2 \Sigma^\dagger) + p.c. \nonumber\\[2mm]
	\CO_{\chi}^{1I} & = &  \Str(F_{1I} \Sigma F_{2I} \Sigma^\dagger) \nonumber\\[2mm]
	\CO_{\chi}^{1T} & = &  \sum_{\mu \neq \nu} \Str(F_{1T} \Sigma F_{2T} \Sigma^\dagger) \nonumber\\
	\CO_{\chi}^{2T} &=&  \sum_{\mu \neq \nu} \Str(\xi_{\mu \nu}  \Sigma F_1 \Sigma^\dagger \xi_{\nu \mu}  \Sigma F_2 \Sigma^\dagger) + p.c. \nonumber\\
	\CO_{\chi}^{3T} &=&  \sum_{\mu \neq \nu} \Str(\xi_{\mu \nu} \Sigma F_1 \Sigma^\dagger)\Str(\xi_{\nu \mu} \Sigma F_2 \Sigma^\dagger) + p.c. \nonumber\\
	\CO_{\chi}^{1V} & = &  \sum_\mu \Big( \Str(F_{1V} \Sigma F_{2V} \Sigma) + \Str(F_{1V} \Sigma^\dagger F_{2V} \Sigma^\dagger) \Big) \nonumber\\
	\CO_{\chi}^{2V} & = &  \sum_\mu \Big( \Str(F_{1V} \Sigma)\Str(F_{2V} \Sigma) + \Str(F_{1V} \Sigma^\dagger)\Str(F_{2V} \Sigma^\dagger) \Big) \nonumber\\
	\CO_{\chi}^{3V} & = &  \sum_\mu \Str(F_{1V} \Sigma)\Str(F_{2V} \Sigma^\dagger) \nonumber\\
	\CO_{\chi}^{1A} & = &  \sum_\mu \Big( \Str(F_{1A} \Sigma F_{2A} \Sigma) + \Str(F_{1A} \Sigma^\dagger F_{2A} \Sigma^\dagger) \Big) \nonumber\\
	\CO_{\chi}^{2A} & = &  \sum_\mu \Big( \Str(F_{1A} \Sigma)\Str(F_{2A} \Sigma) + \Str(F_{1A} \Sigma^\dagger)\Str(F_{2A} \Sigma^\dagger) \Big) \nonumber\\
	\CO_{\chi}^{3A} & = &  \sum_\mu \Str(F_{1A} \Sigma)\Str(F_{2A} \Sigma^\dagger).	
\label{eq:OpList}\end{eqnarray}
Appendix~\ref{app:ops} explains the details of our operator notation, but we note here the important features.  In general, each operator contains an $F_1$ and an $F_2$, which are taste matrices with different, nontrivial \emph{flavor} structures such that $\CO_{\chi}^{K}$ produces the desired kaons at tree-level.  In particular, $F_1$ and $F_2$ with no additional labels are taste $\xi_5$, whereas the remaining $F$'$s$ have ``incorrect'' tastes, e.g. $F_{1T}$ is taste $\xi_{\mu\nu}$.  The fact that many operators with tastes other than $P$ contribute to \bk illustrates the importance of incorporating taste-symmetry breaking when calculating quantities with staggered fermions on the lattice.  
   
\begin{figure}
\begin{tabular}{ccccc}
	\epsfxsize=1.4in \epsffile{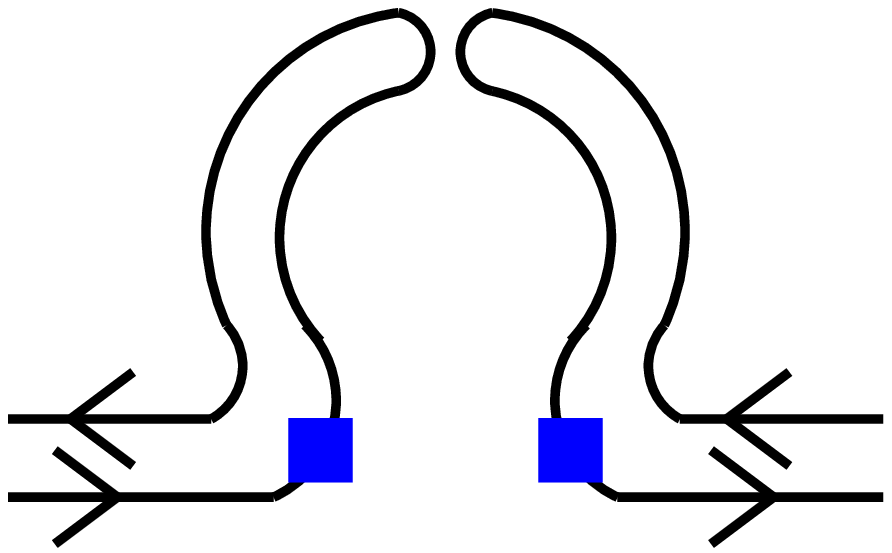} & $\;\;\;\;\;\;$ & \epsfxsize=1.4in \epsffile{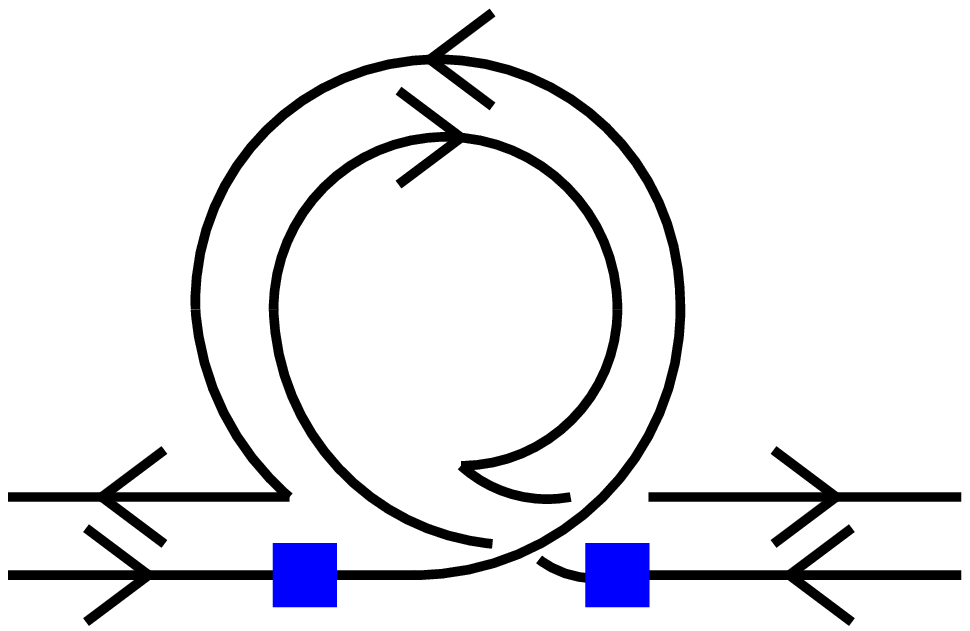} & $\;\;\;\;\;\;\;\;$ & \epsfxsize=1.4in \epsffile{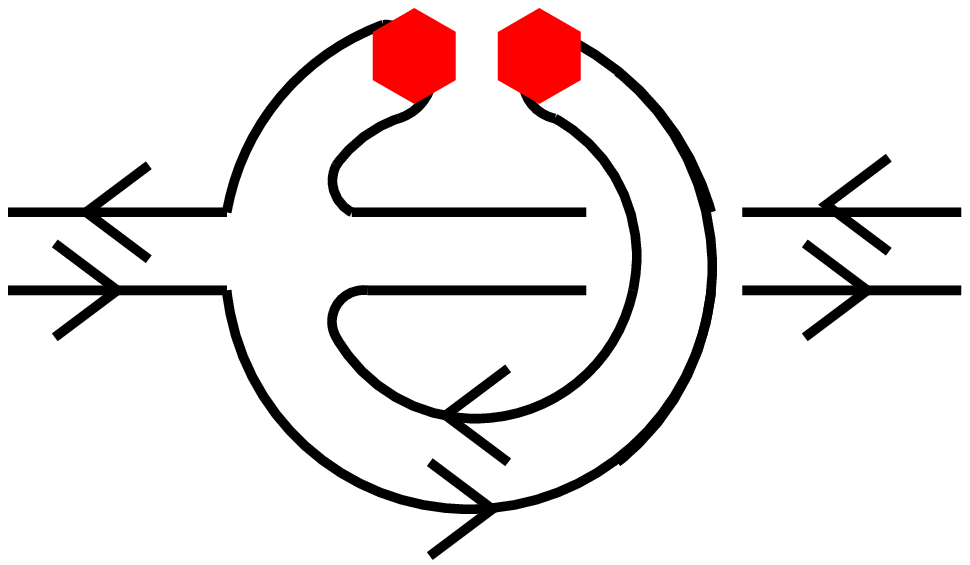} \\
	(a) & $\;\;\;\;\;\;$ & (b) & $\;\;\;\;\;\;$ & (c) \\
	\end{tabular}
	\caption{Quark line diagram contributions to $B_K$ at 1-loop.  One external meson is a $\bar{K^0_1}$ and the other is a $K^0_2$.  The two boxes (hexagons) represent an insertion of the $B_K$ operator.  Each box ``changes'' the valence quark flavor from ($d1 \leftrightarrow s1$) or ($d2 \leftrightarrow s2$).  Each hexagon ``changes'' the valence quark flavor from ($d1 \leftrightarrow s2$) or ($d2 \leftrightarrow s1$). Diagrams (a) and (b) contribute to Figure \ref{fig:BK1Loop}(d) while diagram (c) contributes to Figure \ref{fig:BK1Loop}(e).}\label{fig:BKQuark}\end{figure}

Figure \ref{fig:BK1Loop}(d), in which one of the many \bk operators produces a four-meson vertex and two of the legs are then contracted to form a loop, corresponds to two quark-level diagrams, shown in Figures~\ref{fig:BKQuark}(a) and (b).  The two boxes in each diagram are similar, but not identical: one changes a $d1$ quark into an $s1$ quark while the other changes a $d2$ into an $s2$.
Diagram (a) \emph{must} be disconnected at the quark level because the two external kaons contain different kinds of valence quarks, so it only receives contributions from double-supertrace operators:\footnote{Double supertrace operators $\CO_{\chi}^{3P}$ and $\CO_{\chi}^{3T}$ cannot contribute to \bk at NLO because the loop mesons must be connected through a hairpin propagator, and can only be tastes $I, V,$ or $A$.  For a review of S$\chi$PT and brief discussion of hairpin (quark disconnected) propagators see Appendix~\ref{app:SChPT}.}
\begin{eqnarray}
	\CM^{(a)} &=& \frac{8 \CC_{\chi}^K}{f^4} \int \frac{d^4 q}{(2\pi)^4} (p^2 + q^2) \bigg\{ D_{xx}^I(q) + D_{yy}^I(q) - 2 D_{xy}^I(q)\bigg\} \nonumber\\
	 &-& \frac{16 (2\CC_{\chi}^{2V} + \CC_{\chi}^{3V})}{f^4} \int \frac{d^4 q}{(2\pi)^4} \bigg\{ D_{xx}^A(q) + D_{yy}^A(q) - 2D_{xy}^A(q)\bigg\} \nonumber\\
	 &-&\frac{16 (2\CC_{\chi}^{2A} + \CC_{\chi}^{3A})}{f^4} \int \frac{d^4 q}{(2\pi)^4} \bigg\{ D_{xx}^V(q) + D_{yy}^V(q) - 2D_{xy}^V(q)\bigg\}.
\label{eq:CMa}\end{eqnarray}
The symbols $D^I$, $D^V$, and $D^A$ represent the taste $I$, $V$, and $A$ disconnected propagators, respectively -- we will show their explicit forms later as needed.  Diagram \ref{fig:BKQuark}(b), on the other hand, is connected and only receives contributions from the seven single-supertrace operators\footnote{Recall that $\CO^K_\chi$ and $\CO^N_\chi$ have the same coefficient, which we call $\CC^K_\chi$.}:  
\begin{eqnarray}
  \CM^{(b)} & = & \sum_{B'}\frac{\CC^K_{\chi}f^{B'}}{f^4} \int \frac{d^4 q}{(2\pi)^4} \bigg\{ \frac{2 q^2}{3} \bigg[ \frac{1}{q^2 + m^2_{xy_{B'}}} \bigg] \;-\; \frac{(q^2 + p^2)}{4} \bigg[\frac{1}{q^2 + m^2_{xx_{B'}}} +\frac{1}{q^2 + m^2_{yy_{B'}}} \bigg] \bigg\} \nonumber\\
  &+&  \sum_{B,B'}\frac{\CC^{1B}_{\chi}g^{B B'}}{f^4} \int \frac{d^4 q}{(2\pi)^4} \bigg\{ \frac{1}{3} \bigg[ \frac{1}{q^2 + m^2_{xy_{B'}}} \bigg] \;-\; \frac{1}{4} \bigg[\frac{1}{q^2 + m^2_{xx_{B'}}} + \frac{1}{q^2 + m^2_{yy_{B'}}} \bigg] \bigg\} \nonumber\\
  &+& \sum_{B,B'}\frac{\CC^{2B}_{\chi}h^{B B'}}{f^4} \int \frac{d^4 q}{(2\pi)^4} \bigg\{ \frac{2}{3} \bigg[\frac{1}{q^2 + m^2_{xy_{B'}}} \bigg]\;-\; \frac{1}{4}\bigg[ \frac{1}{q^2 + m^2_{xx_{B'}}} + \frac{1}{q^2 + m^2_{yy_{B'}}} \bigg] \bigg\},
\label{eq:CMb}\end{eqnarray}
where the coefficients $f^{B'}$, $g^{BB'}$, and $h^{BB'}$ are matrices in which $B = I,P,V,A,T$ labels the contributing operator and $B' = I,P,V,A,T$ indicates the loop meson taste:
\begin{eqnarray}
  f^{ B'} =  -8  \left( \begin{array}{ccccc} 1 & 1 & 4 & 4 & 6 \\*  \end{array} \right),
\label{eq:f}\end{eqnarray}
\begin{eqnarray}
  g^{B B'} =  -4 \left( \begin{array}{ccccc} 1 & 1 & -4 & -4 & 6 \\*
	  1 & 1 & 4 & 4 & 6 \\* 
	  -8 & 8 & -16 & 16 & 0 \\* 
	-8 & 8 & 16 & -16 & 0 \\*
	12 & 12 & 0 & 0 & -24 \end{array} \right),
\label{eq:g}\end{eqnarray}
\begin{eqnarray}
  h^{B B'} = 128  \left( \begin{array}{ccccc} 0\;\; & 0\;\; & 0\;\; & 0\;\; & 0\;\; \\*
	  0\;\; & 0\;\;  & 1\;\;  & 1\;\;  & 0\;\;  \\* 
	  0\;\; & 0\;\;  & 0\;\;  & 0\;\;  & 0\;\;  \\* 
	  0\;\; & 0\;\;  & 0\;\;  & 0\;\;  & 0\;\; \\*
	  0\;\; & 0\;\;  & 6\;\;  & 6\;\;  & 12 \end{array} \right).
\label{eq:h}\end{eqnarray}
The sum over $B'$ accounts for how many times a particular loop meson taste contributes to the above expression with a given coefficient -- this is important because a staggered meson's mass depends on its taste as well as its flavor.\footnote{See Ref.~\cite{BA1} for the staggered meson masses at tree-level.}  The pattern $\{1,1,4,4,6\}$ in Eq.~(\ref{eq:f}) counts the number of lattice PGBs with each taste -- thus the coefficient $f^{B'}$ averages over all PGB tastes.  Note that the rows of zeros in Eq.~(\ref{eq:h}) have no deep physical meaning, but simply correspond to the absence of certain operators, e.g. $h^{0B} = (0,0,0,0,0)$ indicates that $\CO^{2I}_\chi$ does not exist.  Nevertheless we include them so as to allow both $B$ and $B'$ to run over the same set of indices.         

Figure~\ref{fig:BK1Loop}(e), in which an operator from the staggered chiral Lagrangian generates the four-meson vertex and one of the \bk operators is inserted in the loop, corresponds to the third quark level diagram, Figure~\ref{fig:BKQuark}(c).  Because the LO staggered chiral Lagrangian contains three types of operators -- the continuum two-derivative term, the continuum mass term, and the $\CO(a^2)$ potential -- there are many contributions to this diagram.  Nevertheless, although the staggered potential contains both single- and double-supertrace operators, we only show a connected vertex in ~\ref{fig:BKQuark}(c).\footnote{The staggered potential is shown explicitly in Appendix~\ref{app:SChPT}.}  This is because it turns out that, for all disconnected 4-PGB vertices from the LO staggered potential in which two of the legs have the desired flavors of external kaons, the various Wick contractions cancel and the vertex is identically zero.  Thus we need only consider connected 4-PGB vertices.  Because the two external kaons have entirely different flavors, they cannot share any quark lines without a \bk operator insertion (box or hexagon), so Figure~\ref{fig:BKQuark}(c) shows the only possible quark flow.  However, it is now easy to see that the \bk operator must change both the quark flavor and the valence set, so we use hexagons rather than boxes to indicate this.  Finally, at the 2-PGB level, double-supertrace \bk operators (by construction) change flavors within a valence set, whereas single-supertrace operators change flavors between sets of valence quarks.  Thus the last diagram only receives contributions from single-supertrace operators:
\begin{eqnarray}
  \CM^{(c)} & = & \sum_{B'}\int \frac{d^4 q}{(2\pi)^4} \bigg[ -\CC^K_{\chi}f^{B'}q^2 + \sum_{B}\Big(\CC_{\chi}^{1B}g^{B B'} -\CC_{\chi}^{2B}h^{B B'}\Big) \bigg] \nonumber\\
	&& \times \bigg[  \frac{ p^2 + q^2 -  2 m^2_{xy_P} + m^2_{xy_{B'}}}{6 f^4}\bigg] \bigg[ \frac{1}{q^2 + m^2_{xy_{B'}}} \bigg]^2.
\label{eq:CMc}\end{eqnarray}
Note that it depends on the \emph{same} coefficients, $f$, $g$, and $h$ as $\CM^{(b)}$ -- this will allow us to combine the two in the following section.

\section{Next-to-Leading Order \bk Results}
\label{sec:Res}

The 1-loop matrix elements determined in the previous section apply to a PQ theory with 4 tastes of each sea quark flavor.  To go from four tastes per sea quark to one taste per sea quark (from the so-called ``4+4+4 theory'' to the ``1+1+1 theory'' when $N_{sea} = 3$), we must multiply every sea quark loop by $1/4$.  At first glance, the quark-level diagrams which contribute to \bk at 1-loop, Figs.~\ref{fig:BKQuark}(a)-(c), do not appear to contain any sea quark loops.  However, sea quark loops implicitly enter the disconnected hairpin propagators when one rediagonalizes the mass matrices of the flavor-neutral, taste $V$, $A$, and $I$, sectors.  In general, the forms of quenched, PQ, and full QCD matrix elements only differ for diagrams with sea quark loops, so the final expression for Eq.~(\ref{eq:CMa}) will change in the three cases, but those for Eqs.~(\ref{eq:CMb}) and (\ref{eq:CMc}) will remain the same.  

\subsection{\bk at NLO in the PQ Theory}

We first consider \bk in the 1+1+1 PQ theory ($m_u \neq m_d \neq m_s$), as it is the most general.  We label the valence quarks by $x$ and $y$, and reserve the labels $u$, $d$, and $s$ for the sea quarks.  Note that the sea quark masses only show up in the PQ results \emph{implicitly} through the masses of the flavor-neutral PGBs: $\pi^0$, $\eta$, and $\eta'$.  Thus the 1+1+1 result can be turned into the 2+1 ($m_u = m_d \neq m_s$) result simply by changing the expressions for the $\pi^0$, $\eta$, and $\eta'$ masses.

The matrix elements that come from \emph{connected} quark-level diagrams are most simple to calculate, so we consider them first.  Neither Fig.\ref{fig:BKQuark}(b) nor Fig.~\ref{fig:BKQuark}(c) contains sea quark loops, so we do not have to insert any factors of $1/4$ by hand -- we need only perform the integrals.  To simplify the resulting expressions, we use a condensed notation for the chiral logarithms.  As in Refs.~\cite{BA1} and \cite{BA2}, we define two functions, $\ell$ and $\tilde \ell$:
\begin{eqnarray}
	\int \frac{d^4 q}{(2\pi)^4} \frac{1}{q^2 + m^2} &\equiv& \frac{1}{16 \pi^2} \ell(m^2)  \\
	\int \frac{d^4 q}{(2\pi)^4} \frac{1}{(q^2 + m^2)^2} &\equiv& \frac{1}{16 \pi^2} \tilde \ell(m^2).  
\end{eqnarray}
The remaining integrals can all be expressed in terms of $\ell$ and $\tilde\ell$:
\begin{eqnarray}
	\int \frac{d^4 q}{(2\pi)^4} \frac{q^2}{q^2 + m^2} &=& \frac{-m^2}{16 \pi^2} \ell(m^2) \\
	\int \frac{d^4 q}{(2\pi)^4} \frac{q^2}{(q^2 + m^2)^2} &=& \frac{1}{16 \pi^2} \bigg( \ell(m^2) - m^2\tilde\ell(m^2) \bigg) \\
	\int \frac{d^4 q}{(2\pi)^4} \frac{q^4}{(q^2 + m^2)^2} &=& \frac{m^2}{16 \pi^2} \bigg(m^2\tilde\ell(m^2) - 2\ell(m^2) \bigg) . 
\end{eqnarray}
Because finite-volume (FV) corrections to \bk only alter the chiral logarithms by turning integrals into sums, one can use either the infinite volume or FV expressions for $\ell$ and $\tilde\ell$ as desired:
\begin{eqnarray}
  \ell(m^2) & = & m^2 \Big( \textrm{ln}\frac{m^2}{\mu_{D.R.}^2} + \delta_1^{FV}(m\textrm{L}) \Big), \;\;\;\;\;\;\;\;\;\, \delta_1^{FV}(m\textrm{L}) = \frac{4}{m\textrm{L}} \sum_{\vec{r}\neq0} \frac{K_1(|\vec{r}|m\textrm{L})}{|\vec{r}|} \\
  \tilde\ell(m^2) & = & -\Big(\textrm{ln}\frac{m^2}{\mu_{D.R.}^2} + 1 \Big) + \delta_3^{FV}(m\textrm{L}), \;\;\;\;\; \delta_3^{FV}(m\textrm{L}) = 2 \sum_{\vec{r}\neq0} K_0(|\vec{r}|m\textrm{L}) ,
\end{eqnarray}
where $\textrm{L}^3$ is the lattice spatial volume and $K_0$ and $K_1$ are Bessel functions of imaginary argument.\footnote{Note that we use dimensional regularization and the same renormalization scheme as Refs.~\cite{BA1, BA2}.}  Thus our final expression for \bk will be completely general.  In terms of $\ell$ and $\tilde\ell$, the 1-loop matrix element contribution to \bk from connected quark-level diagrams, $\CM^{(b)} + \CM^{(c)}$, is   
\begin{eqnarray}
  \CM_{conn}  &=& \frac{1}{64 \pi^2 f^4} \sum_{B'} \Bigg\{ \CC^K_{\chi}f^{B'} \bigg[ m^2_{X_{B'}} \ell(m^2_{X_{B'}}) + m^2_{Y_{B'}}\ell(m^2_{Y_{B'}}) - 2m^2_{xy_{B'}}\ell(m^2_{xy_{B'}}) \bigg] \nonumber\\
	&-&  \sum_B \Big( \CC^{1B}_{\chi}g^{BB'} + \CC^{2B}_{\chi}h^{BB'} \Big)\bigg[ \ell(m^2_{X_{B'}}) + \ell(m^2_{Y_{B'}}) - 2\ell(m^2_{xy_{B'}}) \bigg] \nonumber\\
    &+& \CC^K_{\chi}f^{B'}m^2_{xy_{P}}\bigg[ \ell(m^2_{X_{B'}}) + \ell(m^2_{Y _{B'}}) + 2\ell(m^2_{xy_{B'}}) - 2 m^2_{xy_{B'}} \tilde\ell(m^2_{xy_{B'}}) \bigg] \nonumber\\
  &-& \sum_B \Big( \CC^{1B}_{\chi}g^{BB'} -\CC^{2B}_{\chi}h^{BB'} \Big) m^2_{xy_{P}}\bigg[2 \tilde\ell(m^2_{xy_{B'}})\bigg] \Bigg\}. 
\label{eq:MConn}\end{eqnarray}
We use a condensed notation for the flavor-neutral mesons, in which $m^2_X \equiv m^2_{xx}$ and so forth.  Note that the first two lines vanish when $m_x = m_y$.  Note also that the second and third terms in $\CM_{conn}$ are proportional the mass of the \emph{external} kaons because we have let $p^2 = - m^2_{xy_P}$.  

Next we consider the matrix element that comes from the \emph{disconnected} diagram in Fig.~\ref{fig:BKQuark}(a), Eq.~(\ref{eq:CMa}), which contains tastes $I$, $V$, and $A$ hairpin propagators.  Because the 2-PGB disconnected vertices are dimension 2, an additional hairpin vertex can be canceled by an additional propagator in the S$\chi$PT power-counting, so one must resum the flavor-neutral propagators with all possible numbers of sea quark loops between the external hairpins.  The staggered, flavor-neutral, full propagators were determined in Ref.~\cite{BA1} using the general method outlined in Appendix A of Ref.~\cite{SS1}, but adding factors of $1/4$ for every sea quark loop in the series.  We will just quote the results.

Generically, hairpin propagators contain poles at the mass eigenstates ($\pi^0$, $\eta$, and $\eta'$) and the flavor eigenstates ($X$ and $Y$).\footnote{The staggered factors of 1/4 shift the mass eigenvalues but do not introduce additional poles.  Expressions for the $\pi^0$, $\eta$, and $\eta'$ masses in a theory with 2+1 sea quarks are given in Ref.~\cite{BA1}.  The generalization to the 1+1+1 case is straightforward.}  These multiple-poles can be rewritten as sums of single poles times their residues, e.g.:
\begin{eqnarray}
	D^{V, PQ}_{xy}(q) &= &-a^2 \delta'_V \frac{1}{(q^2 + m^2_{X_V})(q^2 + m^2_{Y_V})} \frac{(q^2 + m^2_{U_V})(q^2 + m^2_{D_V})(q^2 + m^2_{S_V})}{(q^2 + m^2_{\pi^0_V})(q^2 + m^2_{\eta_V})(q^2 + m^2_{\eta'_V})} \label{eq:DVPQ}\\
	& = & -a^2 \delta'_V \!\!\!\!\!\!\!\!\sum_{j = X, Y, \pi^0, \eta, \eta'}\!\frac{1}{q^2 + m^2_{j_V}}R^{[5,3]}_{j_V}(\{ m_X, m_Y, m_{\pi^0}, m_\eta, m_{\eta'} \};\{ m_U, m_D, m_S \}), 
\label{eq:R53}\end{eqnarray} 
where $R^{[5,3]}_{j_V}$ is the residue of the single pole at $q^2 = - m^2_{j_V}$, e.g.:
\begin{equation}
  R^{[5,3]}_{X_V} =  \frac{(m^2_{U_V} - m^2_{X_V})(m^2_{D_V}- m^2_{X_V})(m^2_{S_V}- m^2_{X_V})}{(m^2_{Y_V} - m^2_{X_V})(m^2_{\pi^0_V}- m^2_{X_V})(m^2_{\eta_V}- m^2_{X_V})(m^2_{\eta'_V}- m^2_{X_V})}.
\end{equation}
We use the notation of Refs.~\cite{BA1} and \cite{BA2}, which is most clearly described in the Appendix of \cite{BA2}.  The expression in Eq.~(\ref{eq:DVPQ}) is valid for 3 sea quarks with any combination of masses; an analogous expression holds for $D^{A, PQ}_{xy}(q)$.  Because the \emph{overall} flavor-taste singlet decouples from the PQ theory, when $m_0 \rightarrow \infty$ the expression for $D^{I, PQ}_{xy}(q)$ simplifies further:  
\begin{eqnarray}
	D^{I, PQ}_{xy}(q) &= & -\frac{4 m_0^2}{3} \frac{1}{(q^2 + m^2_{X_I})(q^2 + m^2_{Y_I})} \frac{(q^2 + m^2_{U_I})(q^2 + m^2_{D_I})(q^2 + m^2_{S_I})}{(q^2 + m^2_{\pi^0_I})(q^2 + m^2_{\eta_I})(q^2 + m^2_{\eta'_I})} \\
	& \xrightarrow[m_0 \rightarrow \infty]{} & -\frac{4}{3} \frac{1}{(q^2 + m^2_{X_I})(q^2 + m^2_{Y_I})} \frac{(q^2 + m^2_{U_I})(q^2 + m^2_{D_I})(q^2 + m^2_{S_I})}{(q^2 + m^2_{\pi^0_I})(q^2 + m^2_{\eta_I})} \\
	& = & -\frac{4}{3} \sum_{j = X, Y, \pi^0, \eta}\frac{1}{q^2 + m^2_{j_I}}R^{[4,3]}_{j_I}(\{ m_X, m_Y, m_{\pi^0}, m_\eta \};\{ m_U, m_D, m_S \}). 
\label{eq:R43}\end{eqnarray} 
Some of the taste $I$ disconnected propagators are multiplied by $q^2$, so the residues of their poles are changed:
\begin{eqnarray}
	q^2 D^{I, PQ}_{xy}(q) & \sim & \frac{4}{3} \sum_{j = X, Y, \pi^0, \eta}\frac{m^2_{j_I}}{q^2 + m^2_{j_I}}R^{[4,3]}_{j_I}(\{ m_X, m_Y, m_{\pi^0}, m_\eta \};\{ m_U, m_D, m_S \}),
\end{eqnarray}
where ``$\sim$'' indicates that this relationship only holds within the integral.  

Using the definitions of the disconnected propagators it can be shown that 
\begin{equation}
	D_{xx} + D_{yy} - 2D_{xy} = (m_X^2 - m_Y^2)^2 \frac{\partial}{\partial m_X^2}\frac{\partial}{\partial m_Y^2} \big\{D_{xy} \big\}.
\end{equation}
This greatly simplifies the 1-loop matrix element contribution to \bk from disconnected quark-level diagrams:
\begin{eqnarray}
	\CM_{disc}^{PQ} & =&  \frac{2 \CC_{\chi}^K}{3 \pi^2 f^4} \big(m^2_{X_I} - m^2_{Y_I}\big)^2  \frac{\partial}{\partial m^2_{X_I}} \frac{\partial}{\partial m^2_{Y_I}} \bigg\{ \sum_{j = X, Y, \pi^0, \eta} \ell(m^2_{j_I}) \big(m^2_{xy_P} + m^2_{j_I}\big) R^{[4,3]}_{j_I} \bigg\} \nonumber\\
	&+& \frac{(2\CC_{\chi}^{2V} + \CC_{\chi}^{3V}) a^2 \delta'_A}{\pi^2 f^4}  \big(m^2_{X_A} - m^2_{Y_A}\big)^2  \frac{\partial }{\partial m^2_{X_A}} \frac{\partial }{\partial m^2_{Y_A}} \bigg\{ \sum_{j = X, Y, \pi^0, \eta, \eta'} \ell(m^2_{j_A}) \, R^{[5,3]}_{j_A} )  \bigg\} \nonumber\\
	&+& \frac{(2\CC_{\chi}^{2A} + \CC_{\chi}^{3A}) a^2 \delta'_V}{\pi^2 f^4}  \big(m^2_{X_V} - m^2_{Y_V}\big)^2  \frac{\partial }{\partial m^2_{X_V}} \frac{\partial }{\partial m^2_{Y_V}} \bigg\{ \sum_{j = X, Y, \pi^0, \eta, \eta'} \ell(m^2_{j_V}) \, R^{[5,3]}_{j_V} \bigg\} ,
\label{eq:MDiscPQ}\end{eqnarray}
where we have set $p^2 = - m^2_{xy_P}$ and the arguments of the $R$'s are as in Eqs~(\ref{eq:R53}) and (\ref{eq:R43}).  This compact notation emphasizes that the disconnected matrix element vanishes quadratically as $(m_x - m_y) \rightarrow 0$.

Finally, \bk at NLO in a PQ theory with 1+1+1 sea quarks is 
\begin{equation}
  B_K^{PQ} = B_0 + \frac{3}{8} \left(\frac{\CM_{conn} + \CM_{disc}^{PQ}}{f_{xy_P}^2 m_{xy_P}^2} + \frac{A}{f_{xy_P}^2} + B\frac{m^2_{xy_P}}{f_{xy_P}^2} + C\frac{(m_x - m_y)^2}{f_{xy_P}^2 m_{xy_P}^2} + D\frac{(m_u + m_d + m_s)}{f_{xy_P}^2}\right),
\label{eq:BKPQ}\end{equation}
where $B_0$ is given in Eq.~(\ref{eq:B0}) and $\CM_{conn}$ and $\CM_{disc}^{PQ}$ are given in Eqs.~(\ref{eq:MConn}) and (\ref{eq:MDiscPQ}).  Here we have reintroduced the analytic terms from Appendix~\ref{app:Anal}.  We have checked explicitly that all renormalization scale dependence in the logarithms can be absorbed by the analytic terms.

\subsection{\bk at NLO in the Quenched Theory}

We now find \bk in the quenched theory.  Because of the absence of sea quark loops, additional factors of $1/4$ to reduce the number of tastes per flavor are unnecessary.  Moreover, the flavor-neutral propagators need not be resummed, so their forms are quite simple, e.g.:
\begin{equation}
	D^{V, Quench}_{xy}(q) = -a^2 \delta'_V \frac{1}{(q^2 + m^2_{X_V})(q^2 + m^2_{Y_V})},
\end{equation}
and likewise for $D^{A, Quench}_{xy}(q)$.  However, the flavor-taste singlet ($\eta'_I$) does not decouple in the quenched theory, so we cannot take $m_0 \rightarrow \infty$: 
\begin{equation}
	D^{I, Quench}_{xy}(q) = -\frac{4}{3} \frac{m_0^2 + \alpha q^2}{(q^2 + m^2_{X_I})(q^2 + m^2_{Y_I})}\,,
\end{equation}
where $\alpha$ is an additional chiral parameter that is only present in the quenched theory, not the strong coupling constant.  As before, the nondegenerate propagators can be further simplified into sums of poles times residues:
\begin{eqnarray}
	D^{V, Quench}_{xy}(q) & = & -a^2 \delta'_V \frac{1}{m^2_{Y_V} - m^2_{X_V}} \bigg[ \frac{1}{q^2 + m^2_{X_V}} - \frac{1}{q^2 + m^2_{Y_V}} \bigg] \nonumber\\
	D^{I, Quench}_{xy}(q) & = & \bigg( \frac{-4}{3}\bigg) \frac{1}{m^2_{Y_I} - m^2_{X_I}} \bigg[ \frac{m_0^2 - \alpha m^2_{X_I}}{(q^2 + m^2_{X_I})} - \frac{m_0^2 - \alpha m^2_{Y_I}}{(q^2 + m^2_{Y_I})} \bigg].
\end{eqnarray}

Because the connected quark-level diagrams that contribute to \bk at 1-loop, Figs.~\ref{fig:BKQuark}(b) and (c), do not contain sea quark loops, the corresponding matrix elements remain unchanged from the PQ theory.  The matrix elements from disconnected quark-level diagrams, after setting $p^2 = -m^2_{xy_P}$, are
\begin{eqnarray}
	\CM_{disc}^{Quench} & =&  \frac{2 \CC_{\chi}^K}{3 \pi^2 f^4} \big(m^2_{X_I} - m^2_{Y_I}\big)^2  \!\frac{\partial}{\partial m^2_{X_I}} \frac{\partial}{\partial m^2_{Y_I}} \bigg\{ \sum_{j = X, Y} \!\!\ell(m^2_{j_I}) \big(m^2_{xy_P} + m^2_{j_I}\big) \big(m_0^2 - \alpha m^2_{j_I}\big) R^{[2,0]}_{j_I} \bigg\} \nonumber\\
	&+& \frac{(2\CC_{\chi}^{2V} + \CC_{\chi}^{3V}) a^2 \delta'_A}{\pi^2 f^4}  \big(m^2_{X_A} - m^2_{Y_A}\big)^2  \frac{\partial}{\partial m^2_{X_A}} \frac{\partial}{\partial m^2_{Y_A}} \bigg\{  \sum_{j = X, Y} \ell(m^2_{j_A}) \, R^{[2,0]}_{j_A} \bigg\} \nonumber\\
	&+& \frac{(2\CC_{\chi}^{2A} + \CC_{\chi}^{3A}) a^2 \delta'_V}{\pi^2 f^4}  \big(m^2_{X_V} - m^2_{Y_V}\big)^2  \frac{\partial}{\partial m^2_{X_V}} \frac{\partial}{\partial m^2_{Y_V}} \bigg\{\sum_{j = X, Y} \ell(m^2_{j_V}) \, R^{[2,0]}_{j_V}  \bigg\} ,
\end{eqnarray}
where 
\begin{equation}
R^{[2,0]}_{X_V} \equiv R^{[2,0]}_{X_V}(\{m_X,m_Y\}) =  \frac{1}{(m^2_{Y_V} - m^2_{X_V})} = - R^{[2,0]}_{Y_V}\,.
\end{equation}

Therefore \bk at NLO in the Quenched theory is
\begin{equation}
  B_K^{Quench} = B_0 + \frac{3}{8} \left(\frac{\CM_{conn} + \CM_{disc}^{Quench}}{f_{xy_P}^2 m_{xy_P}^2} + \frac{A}{f_{xy_P}^2} + B\frac{m^2_{xy_P}}{f_{xy_P}^2} + C\frac{(m_x - m_y)^2}{f_{xy_P}^2 m_{xy_P}^2}\right), 
\end{equation}
where $B_0$ is given in Eq.~(\ref{eq:B0}) and $\CM_{conn}$ is  given in Eq.~(\ref{eq:MConn}).  We emphasize that the undetermined coefficients, $\CC^i_\chi$ and $A$-$C$, are \emph{different} in the quenched theory than in QCD, and that there is no $D$ term.  

\subsection{\bk at NLO in the Full (2+1) Theory}

Finally we determine \bk in full QCD, i.e. $m_x = m_d$ and $m_y = m_s$, with 2+1 sea quarks ($m_u = m_d \neq m_s$).  Although one can, in principle, just take the limit of the final PQ expression, Eq.~(\ref{eq:BKPQ}), it is easier to start from the expressions for the 1-loop matrix elements in terms of integrals.  

Because QCD is a physical theory, the disconnected propagators only have single poles.  It is easy to show this explicitly for the 2+1 theory, in which  $m^2_{\pi^0} = m^2_U = m^2_D$ and parts of the numerator and denominator cancel.  For example, the taste $V$ (and $A$) disconnected propagators are:
\begin{eqnarray}
	D^{V, \:2+1}_{dd}(q) &=& -a^2 \delta'_V \bigg[ \frac{1}{q^2 + m^2_{D_V}} \bigg( \frac{(m^2_{S_V} - m^2_{D_V})}{(m^2_{\eta_V} - m^2_{D_V})(m^2_{\eta'_V} - m^2_{D_V})} \bigg) + (D \rightarrow \eta \rightarrow \eta' \rightarrow D) \bigg] \\
	D^{V, \:2+1}_{ss}(q) &=& -a^2 \delta'_V \bigg[ \frac{1}{q^2 + m^2_{S_V}} \bigg( \frac{(m^2_{D_V} - m^2_{S_V})}{(m^2_{\eta_V} - m^2_{S_V})(m^2_{\eta'_V} - m^2_{S_V})} \bigg) + (S \rightarrow \eta \rightarrow \eta' \rightarrow S) \bigg] \\
	D^{V, \:2+1}_{ds}(q) &=& -a^2 \delta'_V \bigg[ \frac{1}{q^2 + m^2_{\eta_V}}\frac{1}{(m^2_{\eta'_V} - m^2_{\eta_V})}  + (\eta \leftrightarrow \eta') \bigg],
\end{eqnarray}
where the arrows indicate all possible permutations of masses.  As in the PQ theory, the taste singlet $\eta'$ decouples, so we can use the explicit large-$m_0$ expressions for the masses,
\begin{eqnarray}
	m^2_{\pi^0_I} & = & m^2_{U_I} = m^2_{D_I}\\
	m^2_{\eta_I} & = & \frac{m^2_{D_I}}{3} + \frac{ 2 m^2_{S_I}}{3} \\
	m^2_{\eta'_I} & = & m_0^2,
\end{eqnarray}
to drastically simplify the taste $I$ disconnected propagators:
\begin{eqnarray}
	D^{I, \:2+1}_{dd}(q) &\xrightarrow[m_0 \rightarrow \infty]{}&  -2 \frac{1}{q^2 + m^2_{D_I}} + \frac{2}{3}\frac{1}{q^2 + m^2_{\eta_I}}  \\
	D^{I, \:2+1}_{ss}(q) &\xrightarrow[m_0 \rightarrow \infty]{}&  -4 \frac{1}{q^2 + m^2_{S_I}} + \frac{8}{3}\frac{1}{q^2 + m^2_{\eta_I}}  \\
	D^{I, \:2+1}_{ds}(q) &\xrightarrow[m_0 \rightarrow \infty]{}&  - \frac{4}{3}\frac{1}{q^2 + m^2_{\eta_I}} .
\end{eqnarray}

In this section we only show the 1-loop matrix elements that come from disconnected quark-level diagrams:
\begin{eqnarray}
	\CM_{disc}^{2+1} &=& \frac{\CC_{\chi}^K}{\pi^2 f^4}\bigg\{ (m^2_{ds_P} + m^2_{D_I})\ell(m^2_{D_I}) +  2 (m^2_{ds_P} + m^2_{S_I})\ell(m^2_{S_I}) - 3(m^2_{ds_P} + m^2_{\eta_I})\ell(m^2_{\eta_I}) \bigg\} \nonumber\nonumber\\
	&-& \frac{(2\CC_{\chi}^{2V} + \CC_{\chi}^{3V})a^2 \delta'_A }{\pi^2 f^4} \bigg\{ 2\bigg[\ell(m^2_{\eta_A})\frac{1}{(m^2_{\eta'_A} - m^2_{\eta_A})}  + (\eta \leftrightarrow \eta') \bigg] \nonumber\\
	&& -\bigg[\ell(m^2_{D_A})\bigg( \frac{(m^2_{S_A} - m^2_{D_A})}{(m^2_{\eta_A} - m^2_{D_A})(m^2_{\eta'_A} - m^2_{D_A})} \bigg) + (D \rightarrow \eta \rightarrow \eta' \rightarrow D) \bigg]\nonumber\\
	&& - \bigg[\ell(m^2_{S_A})\bigg( \frac{(m^2_{D_A} - m^2_{S_A})}{(m^2_{\eta_A} - m^2_{S_A})(m^2_{\eta'_A} - m^2_{S_A})} \bigg) + (S \rightarrow \eta \rightarrow \eta' \rightarrow S) \bigg] \bigg\} \nonumber\\
	&-& \frac{(2\CC_{\chi}^{2A} + \CC_{\chi}^{3A})a^2 \delta'_V }{\pi^2 f^4}\bigg\{ 2\bigg[\ell(m^2_{\eta_V}) \frac{1}{(m^2_{\eta'_V} - m^2_{\eta_V})}  + (\eta \leftrightarrow \eta') \bigg] \nonumber\\
	&& -\bigg[\ell(m^2_{D_V}) \bigg( \frac{(m^2_{S_V} - m^2_{D_V})}{(m^2_{\eta_V} - m^2_{D_V})(m^2_{\eta'_V} - m^2_{D_V})} \bigg) + (D \rightarrow \eta \rightarrow \eta' \rightarrow D) \bigg]\nonumber\\
	&& - \bigg[\ell(m^2_{S_V}) \bigg( \frac{(m^2_{D_V} - m^2_{S_V})}{(m^2_{\eta_V} - m^2_{S_V})(m^2_{\eta'_V} - m^2_{S_V})} \bigg) + (S \rightarrow \eta \rightarrow \eta' \rightarrow S) \bigg] \bigg\}.
\end{eqnarray}
The matrix elements which come from connected diagrams are \emph{minimally} changed from the PQ ones in that $x \rightarrow d$ and $y \rightarrow s$.  Therefore \bk at NLO in full QCD with 2+1 sea quarks is 
\begin{eqnarray}
  B_K^{2+1} &=& B_0 + \frac{3}{8} \left(\frac{\CM_{conn}(x \rightarrow d, y \rightarrow s) + \CM_{disc}^{2+1}}{f_{ds_P}^2 m_{ds_P}^2}\right) \nonumber\\
	&+& \frac{3}{8} \left(\frac{A}{f_{ds_P}^2} + B\frac{m^2_{ds_P}}{f_{ds_P}^2} 
	+  C\frac{(m_d - m_s)^2}{f_{ds_P}^2 m^2_{ds_P}} + D\frac{(2m_d + m_s)}{f_{ds_P}^2}\right),
\end{eqnarray}
where $B_0$ is given in Eq.~(\ref{eq:B0}) and $\CM_{conn}$ and is  given in Eq.~(\ref{eq:MConn}). 

\section{Guide for Lattice Determination of \bk}
\label{sec:Lat}

Any lattice determination of \bk necessarily entails two steps: first, fit the $\chi$PT expression to the lattice data and determine the unknown coefficients, and, second, use the resulting function to extrapolate \bk to its value at the physical quark masses in the continuum limit.  The large number of undetermined constants in the staggered $\chi$PT expression for \bk at NLO could make the fitting procedure quite difficult, and certainly necessitates a lot of data.  Our goal is therefore is to find alternative ways to access some of these coefficients.  Once measured, they can be used in the full NLO S$\chi$PT \bk fit.  

We begin by reviewing explicitly all of the parameters in the expression for \bk at NLO.  We presume that \bk will be fit as a function of the kaon mass squared, so we do not consider the latter to be a fit parameter.  In order to properly count the number of undetermined coefficients, we must combine the PQ expression for $B_K$, Eq.~(\ref{eq:BKPQ}), with Tables~\ref{tab:NLOCoeff} and~\ref{tab:AnalCoeff}.  Only thirteen of the fifteen operators listed in Eq.~(\ref{eq:OpList}) in fact appear in Eq.~(\ref{eq:BKPQ}).  Operators $\CO_{\chi}^{3P}$ and $\CO_{\chi}^{3T}$ turn out not to contribute to \bk at NLO.  Operators $\CO_\chi^{2V}$ and $\CO_\chi^{3V}$ (and similarly $\CO_\chi^{2A}$ and $\CO_\chi^{3A}$) turn out to enter the expression for \bk as a single linear combination, and can therefore be associated with a single coefficient.  Thus we can see from  Table~\ref{tab:NLOCoeff} that there are 23 coefficients associated with NLO chiral operators, in addition to $\CC^K_\chi$.  Similarly, Table~\ref{tab:AnalCoeff} shows that there are 6 linearly-independent analytic terms.  Finally, certain parameters appear in all S$\chi$PT calculations:  the four tree-level mass splittings, the two hairpin parameters, and $f_{xy_P}$.  Thus the expression for \bk at NLO is given in terms of 37 undetermined coefficients.   Fortunately, some of these parameters have already been determined on the MILC 2+1 dynamical field configurations \cite{MILCf}.  In particular, the taste Goldstone meson decay constant, as well as the mass splittings between the other PGB tastes, are already known.\footnote{We note that the vector hairpin is poorly determined from the MILC data, and that effects of the vector and axial vector hairpins are, in practice, difficult to disentangle in S$\chi$PT fits.}  However, this still leaves \emph{32 undetermined fit parameters}.

An obvious way to reduce the number of parameters is to perform a fit to a single lattice spacing.  This reduces the coefficients in Table~\ref{tab:NLOCoeff} to one per operator (or per linear combination), i.e. from 23 to 9, and the analytic terms in Table~\ref{tab:AnalCoeff} from 6 to 4.  Clearly one must eventually fit to multiple lattice spacings.  We suggest, however, that is preferable to fit \bk in a multi-stage process.  First perform separate NLO \bk fits at multiple lattice spacings with this reduced set of parameters.  As we will show, a single lattice spacing fit of $B_K$, in combination with tree-level fits of other matrix elements, can be used to separate \emph{almost} all of the perturbative and discretization errors from the desired continuum result.  Second remove these errors at each lattice spacing.  Third fit the results after error subtraction simultaneously to multiple lattice spacings using the appropriately simplified S$\chi$PT expression.  Last send the remaining errors, which have now been obtained, to zero, and calculate a value for the continuum parameter $\hat{B}_K$.\footnote{The explicit expression for $\hat{B}_K$ in terms of ${B}_K$ is given in Eq.~(\ref{eq:BKHat}).}

\bigskip

We now discuss step one: fitting the NLO expression for \bk at a single lattice spacing.  Because this still involves 16 parameters, it is qualitatively useful to first consider the case of degenerate valence quarks, i.e. $m_x = m_y$. Recall from the previous sections that the NLO expression for \bk greatly simplifies in this limit.  The matrix element from the disconnected diagram vanishes, thereby removing coefficients $\CC_\chi^{2V}$ -- $\CC^{3A}_\chi$ and the hairpin parameters, as do half of the terms in the matrix element from the connected diagrams.  The degenerate mass case is therefore the appropriate testing-ground for whether the NLO expression describes the data at all, i.e. if one is in the chiral regime, before moving on to the more complex, but physical, situation of $m_x \neq m_y$.

In fact, the degenerate mass case is also quantitatively useful.  Recall that the dominant contribution to \bk at NLO comes from the tree-level matrix element of $\CO_\chi^K$, and is proportional to $\CC^K_\chi$: 
\begin{equation}
	\frac{\langle\bar{K^0_1}_P \,|\,  \CO_{1-loop}(\textrm{taste P}) \,|\, {K^0_2}_P\rangle}{\CM_\textrm{vac}}  =  6\frac{\CC_{\chi}^{K}}{f^4} + \ldots\,.
\end{equation}
Because we wish to determine \bk (and consequently $\CC^K_\chi$) to NLO accuracy, the above expression is insufficient.\footnote{In fact, at first glance, we also need $f$ at NLO.  However, this is really unnecessary because $\CC_\chi^K$ enters both the LO and NLO contributions to \bk as the combination $\CC_\chi^K/f^4$.  Thus we can treat the quantity $(\CC_\chi^K/f^4)$ as the undetermined fit parameter, and likewise for the other NLO coefficients.}  An NLO fit to \bk with degenerate valence quark masses, however, can potentially be used to determine $\CC^K_\chi$ to NLO accuracy.  In this limit, \bk at NLO has the following form:  
\begin{eqnarray}
  B_K &=& 12 \frac{\CC_{\chi}^{K}}{f^4}\left\{1 + \frac{1}{512 \pi^2 f_{xy_P}^2} \sum_{B'} f^{B'} \bigg( \ell(m^2_{K_{B'}}) - \frac{1}{2} m^2_{K_{B'}} \tilde\ell(m^2_{K_{B'}}) \bigg) \right\} \nonumber\\
  &-& \frac{3}{256 \pi^2 f_{xy_P}^2} \sum_{B'} \tilde\ell(m^2_{K_{B'}}) \sum_B \left(\frac{\CC^{1B}_\chi}{f^4} g^{BB'} - \frac{\CC^{2B}_\chi}{f^4} h^{BB'} \right)  \nonumber\\
  &+& A\frac{3}{8 f_{xy_P}^2} + B\frac{3m^2_{xy_P}}{8f_{xy_P}^2} + D\frac{3(m_u + m_d + m_s)}{8f_{xy_P}^2}\,.
\label{eq:BKSimp}\end{eqnarray}
It depends on $\CC_\chi^K$, now at NLO, three analytic terms, and five linear combinations of other operator coefficients, $\sum_B (\CC^{1B}_\chi g^{BB'} - \CC^{2B}_\chi h^{BB'})$ for $B' = I,P,V,A,T$.  As we show below, these five combinations can be determined to the required accuracy using other matrix elements.  Once they are found, we would like to use a fit of Eq.~(\ref{eq:BKSimp}) to extract $\CC^K_\chi$ to NLO accuracy.  Unfortunately, this turns out not to be possible;  separation of the NLO part of  $\CC^K_\chi$ from the analytic term $A$ requires a simultaneous fit to multiple lattice spacings.  We can, however, consider their sum, $(\CC^K_\chi/f^4 + 3A/16 f_{xy_P}^2)$, to be a single fit parameter that \emph{can} be determined NLO accuracy.  We can also determine the remaining analytic terms, $B$ and $D$, using Eq.~(\ref{eq:BKSimp}) at a single lattice spacing.  

The other simple handle that one has in lattice simulations, besides the quark masses, is the flavors and tastes of the external states.  At tree-level, double supertrace operators have nonvanishing matrix elements between a $\bar{K^0_1}$ ($s1 \bar{d1}$) and a $K^0_2$ ($\bar{s2} d2$).  In contrast, single supertrace operators have nonvanishing matrix elements between ``mixed'' kaons:  $s1 \bar{d2}$ and $\bar{s2} d1$.  Both single and double supertrace operators can have nonvanishing matrix elements between all five possible tastes:  $I, P, V, A,$ and $T$.  Thus, by evaluating the matrix element of $\CO_{1-loop}(\textrm{taste P})$ between different external states, we can naively hope to determine as many as nine independent coefficients for use in the larger NLO S$\chi$PT fit.  In essence, while the two sets of valence quarks and multiple tastes introduce many new operators into staggered $\chi$PT, they also allow the independent determination of more operator coefficients. 

First consider the double supertrace operators, of which there are four:  $\CO_\chi^{2V}$, $\CO_\chi^{3V}$, $\CO_\chi^{2A}$, and $\CO_\chi^{3A}$.  Clearly, to extract the coefficient of $\CO_\chi^{2V}$ (or $\CO_\chi^{2A}$) we need to choose taste vector (or axial) kaons for our external states.  The resulting matrix elements are:
\begin{eqnarray}
	\langle\bar{K^0_1}_{V} | \CO_{1-loop}(\textrm{taste P}) | {K^0_2}_{V}\rangle &=& \frac{-16}{f^2}(2\CC^{2V}_\chi - \CC^{3V}_\chi) \\
	\langle\bar{K^0_1}_{A} | \CO_{1-loop}(\textrm{taste P}) | {K^0_2}_{A}\rangle &=& \frac{-16}{f^2}(2\CC^{2A}_\chi - \CC^{3A}_\chi),
\end{eqnarray}
so we can only determine the linear combination of coefficients $(2\CC^{2V}_\chi - \CC^{3V}_\chi)$.  However, in the 1-loop contributions to $B_K$, these coefficients enter in a different linear combination: $(2\CC^{2V}_\chi + \CC^{3V}_\chi)$.  Thus we cannot in any simple way separately extract the coefficients of the two supertrace operators.  Nevertheless, calculation of these matrix elements should still be done, as it provides an estimate of the size of the NLO coefficients.  Moreover, because the matrix element of $\CO_{1-loop}(\textrm{taste P})$ between taste $I$ and $T$ kaons should be zero at this order:
\begin{eqnarray}
	\langle\bar{K^0_1}_{I} | \CO_{1-loop}(\textrm{taste P}) | {K^0_2}_{I}\rangle &=& 0\; (+\, \mbox{NNLO}) \\
	\langle\bar{K^0_1}_{T} | \CO_{1-loop}(\textrm{taste P}) | {K^0_2}_{T}\rangle &=& 0\; (+\, \mbox{NNLO})\,,
\end{eqnarray}
they can be used to gauge the size of the higher-order terms, thereby testing the validity of our phenomenological power-counting scheme and determining whether or not it is necessary, as is done in Ref.~\cite{MILCf}, to introduce a small subset of NNLO terms.

Next consider the single supertrace operators.  Extracting many of their coefficients turns out to be feasible, although not completely trivial.  There are eight such operators -- $\CO_\chi^N$, $\CO_\chi^{1P}$, $\CO_\chi^{2P}$, $\CO_\chi^{1T}$, $\CO_\chi^{1I}$, $\CO_\chi^{2T}$, $\CO_\chi^{1V}$, and $\CO_\chi^{1A}$ -- and five possible matrix elements. We can combine them into a single equation using the same coefficients as in Eq.~(\ref{eq:BKSimp}):
\begin{eqnarray}
  	\frac{\langle\bar{K^0_1}^{mix}_{B'} | \CO_{1-loop}(\textrm{taste P}) | {K^0_2}^{mix}_{B'}\rangle}{\CM_\textrm{vac}} &=& \frac{3}{8}\frac{1}{N^{B'}f^4 m_{K_{P}}^2}\Bigg[ \CC^K_\chi f^{B'} m_{K_{B'}}^2 \nonumber\\
	&& + \sum_B\bigg(\CC^{1B}_\chi g^{BB'} - \CC^{2B}h^{BB'}\bigg)\Bigg]\label{eq:MatTree},
\end{eqnarray}
where
\begin{eqnarray}
  N^{ B'} =  \left( \begin{array}{ccccc} -1 & -1 & 4 & 4 & -6 \\*  \end{array} \right)\,.
\label{eq:N}\end{eqnarray}
The coefficient $\CC_\chi^K$ can be separated through its quark mass dependence, but this is not useful because it is only determined at LO in this tree-level matrix element.  However, the numerical values of the five linear combinations of coefficients corresponding to $B' = I,P,V,A,T$ are needed, along with the degenerate mass \bk result, Eq.~(\ref{eq:BKSimp}), to determine the combination of coefficients $(12\CC^K_\chi/f^4 + 3A/8 f_{xy_P}^2)$ to NLO accuracy.  The five matrix elements in Eq.~(\ref{eq:MatTree}) also aid in the true, $m_x \neq m_y$, \bk determination because we can solve for five coefficients in terms of the remaining two, and substitute the resulting functions into the NLO S$\chi$PT expression before fitting.  

Thus, after exploiting both degenerate mass and mixed-flavor matrix elements, we are left with an NLO fit at a single lattice spacing to only \emph{seven parameters}:  $\delta'_V$, $\delta'_A$, $(2\CC^{2V}_\chi + \CC^{3V}_\chi)$, $(2\CC^{2A}_\chi + \CC^{3A}_\chi)$, $C$, and two additional NLO coefficients of our choice from Eq.~(\ref{eq:MatTree}).  This should certainly be feasible.  Once this is done, the next step is to remove as much of the discretization and perturbative error as possible \emph{before} combining the data from multiple lattice spacings.  We do this by setting all coefficients that are multiplied by $a^2$, $a_\alpha^2$, $\alpha/4\pi$, and $\alpha^2$ to zero.\footnote{This is obviously the correct way to remove discretization errors.  The right way to remove perturbative errors, however, is less apparent, since we have only \emph{partially} matched the continuum operator at 1-loop.  We will later show that it is, in fact, correct to simply set their coefficients to zero.}  One can see now why it is preferable to fit to one lattice spacing at a time -- it avoids having to give each operator in Table~\ref{tab:NLOCoeff} a separate coefficient for each expansion parameter.  Doing so would clearly be a waste of effort, since one only needs the total error, not the individual coefficients, in order to remove it.  We emphasize, however, that there is one coefficient that cannot be removed at this stage -- $A$.  It is tied up in the linear combination $(12\CC^K_\chi/f^4 + 3A/8 f_{xy_P}^2)$.  We must therefore remove all of the coefficients of operators in Table~\ref{tab:NLOCoeff} and then perform a new fit to multiple lattice spacings.\footnote{Strictly speaking we should fit to the quantity $\hat{B}_K$, which is independent of lattice spacing, as we explain below.}  Such a fit will only contain the continuum parameter $\CC^K_\chi$ and $A$, the latter of which must now be separated into three pieces of $\CO(a^2)$, $\CO(a_\alpha^2)$, and $\CO(\alpha^2)$.  It therefore has \emph{four parameters}, which is a clear improvement over the 37 parameters that would have originally been needed in a fit to multiple lattice spacings.  Once we have separated the three contributions to $A$ from the coefficient $\CC^K_\chi$ at NLO, we will remove them as well.   

\bigskip

Finally, assuming that we have successfully carried out the fitting procedure outlined above, we are ready to turn the result into a continuum value for $\hat{B}_K$.  We therefore review the procedure for matching lattice and continuum operators.  Along the way, we clarify some of the points and justify some of the assumptions made previously.  

After removal of discretization errors, multiple lattice regularized operators match onto a given continuum operator in the following manner:
\begin{equation}
  \CO^\textrm{cont}(\mu) = Z(\mu,a) \CO^\textrm{lat}(a)\,,
\label{eq:OpMatch}\end{equation}
where $\CO^\textrm{cont}(\mu)$ is a single operator, $\CO^\textrm{lat}(a)$ is a vector of operators, and the vector $Z$ relates the two.  The 1-loop expression for $Z(\mu,a)$ is typically written in the following way:
\begin{equation}
	Z(\mu,a) = 1 + \frac{\alpha}{4\pi}c(\mu a)\,,
\label{eq:Zsimp}\end{equation}
where $c$ is a vector that contains the 1-loop matching coefficients.  However, Eq.~(\ref{eq:Zsimp}) clearly has a problem:  what is the appropriate scale at which to evaluate $\alpha$?  This ambiguity can be avoided by using the exact perturbative formula for $Z$ given by Ji in Ref~\cite{Ji:1995vv}:
\begin{equation}
  Z(\mu,a) = \textrm{exp}\Bigg(-\int_0^{g(\mu)} dg'' \frac{\gamma_\textrm{cont}(g'')}{\beta_\textrm{cont}(g'')} \Bigg)\, T_{g'}\,\textrm{exp}\Bigg(-\int_{g(a)}^0 dg' \frac{\gamma_\textrm{lat}(g')}{\beta_\textrm{lat}(g')} \Bigg)\,,
\label{eq:Zmua}
\end{equation}
where $T_{g'}$ indicates a $g'$-ordered product\footnote{Note that there is no $T_{g''}$ because we are considering the case in which there is only one continuum operator and no mixing on the continuum side.} and $\gamma$ and $\beta$ are the anomalous dimension and beta function, respectively.\footnote{Note also that the coupling $g(a)$ can be chosen to be a continuum-like coupling such as $\bar{\textrm{MS}}$ \cite{Gupta:1996yt}.}  Equation~(\ref{eq:Zmua}) is easy to understand physically; it shows that matching is a three-step process.  First run the lattice operators calculated at a scale $q^* = 1/a$ up to $a=0$.  Next match the lattice operators at this scale onto the continuum operator at a scale $\mu=\infty$.  Finally run the continuum operator down to a scale $\mu$, which is typically $2 \GeV$.  This procedure cleanly separates the lattice scale, $a$, from the continuum scale, $\mu$.  

Mathematically, Eq.~(\ref{eq:Zmua}) can be used to derive the following relationship between the continuum and lattice operators:
\begin{eqnarray}
	&& \alpha(\mu)^{-\gamma_0/2\beta_0}  \bigg[1 + \frac{\alpha(\mu)}{4\pi}J_\textrm{cont} + \CO(\alpha^2) \bigg] \CO^\textrm{cont}(\mu) = \nonumber\\
	&& \alpha(q^*)^{-\gamma_0/2\beta_0}\, \bigg[1 + \frac{\alpha(q^*)}{4\pi}J_\textrm{cont} + \CO(\alpha^2)\bigg]\bigg[1 + \frac{\alpha(q^*)}{4\pi}c(q^*a) + \CO(\alpha^2) \bigg] \CO^\textrm{lat}(a)\,,
\label{eq:Zhor}\end{eqnarray}
where
\begin{equation}
	J = \frac{\beta_1 \gamma_0}{2\beta_0^2} - \frac{\gamma_1 }{2\beta_0}\,,
\end{equation}
and the vector $c$ contains the standard 1-loop matching coefficients.  This formulation of perturbative matching is extremely useful.  Because the matching coefficients only appear on the ``lattice side'' of the equation, and no longer depend on the continuum scale $\mu$, one can choose any reasonable scale at which to calculate the matching coefficients -- a natural choice is $q^*=1/a$.  One must, of course, then evaluate all $\alpha$'s in the NLO S$\chi$PT expression for \bk that arise due to truncation of perturbative matching factors at this same scale.  

Equation~(\ref{eq:Zhor}) also justifies our treatment of errors associated with perturbative matching.  Recall that the standard matching procedure is to \emph{only match to taste P operators} at 1-loop.  In terms of the above matching expression, this means that one calculates the matrix element of $[1 + c(\alpha/4\pi)]\CO^\textrm{lat}_P$, where the vector $\CO^{lat}_P$ contains only taste $P$ operators.  However, this is not correct, as lattice operators of \emph{all tastes} are required to correctly match the continuum operator.  We account for this error by including ``wrong-taste'' operators in the S$\chi$PT fit with coefficients of $\CO(\alpha/4\pi)$.  Conversely, if we \emph{were} to correctly match onto the continuum \bk operator at 1-loop, we would \emph{not need} to include such operators.  Thus we conclude that in order to extract the correct result for \bk in the continuum we must set the coefficients of $\CO(\alpha/4\pi)$ in Table~\ref{tab:NLOCoeff} to zero.  One can use the same logic to show that \emph{all} of the perturbative matching errors should be dealt with in this manner, i.e. we must also set the $\CO(\alpha^2)$ coefficients in Tables~\ref{tab:NLOCoeff} and \ref{tab:AnalCoeff} to zero.  This result is fortunate, as it allows treatment of all errors -- both from discretization and perturbative matching -- in the same, simple manner.  

Finally, Eq.~(\ref{eq:Zhor}) shows that, once one has successfully fit the S$\chi$PT expression for \bk and removed the discretization and perturbative matching errors at each lattice spacing, constructing the renormalization group invariant quantity, $\hat{B}_K$ is trivial.  This is because, to 1-loop order \cite{Crisafulli:1995ad}
:
\begin{equation}
   \hat{B}_K = \alpha(\mu)^{-\gamma_0/2\beta_0}  \bigg[1 + \frac{\alpha(\mu)}{4\pi}J_\textrm{cont} \bigg] B^{\bar{\textrm{MS}}}_K(\mu)\,,
\label{eq:BKHat}\end{equation} 
so
\begin{equation}
	\hat{B}_K = \alpha(q^*)^{-\gamma_0/2\beta_0}\, \bigg[1 + \frac{\alpha(q^*)}{4\pi}J_\textrm{cont} \bigg] B_K^\textrm{lat,subtracted}\,,
\label{eq:BKform}\end{equation}
where $B_K^\textrm{lat,subtracted}$ is the lattice value for \bk at the scale $q^*$ with  discretization and perturbative matching errors subtracted.  Conversely, we can use Eq.~(\ref{eq:BKform}) to remove the remaining errors from $\hat{B}_K$ in the following way.  Once we have values for \bk with all errors but those from $A$ removed at each lattice spacing, we turn them into values for $\hat{B}_K$ by multiplying by $\alpha(q^*)^{-\gamma_0/2\beta_0}\, [1 + J_\textrm{cont} \alpha(q^*)/4\pi]$.  We then simultaneously fit $\hat{B}_K$ at multiple lattice spacings to separate the continuum parameter $\CC^K_\chi$ from $A$.  Next we subtract $A$ to get a continuum value for $\hat{B}_K$.  Last we extrapolate the continuum result to the physical quark masses using the continuum $\chi$PT expression.     

\bigskip

Even with the implementation of all of these suggestions, fitting the staggered \bk data will be highly nontrivial.  One would clearly like to somehow decrease the number of operators, or at least undetermined coefficients, that contribute to \bk at NLO.  One way to do this is by using improved links.  This drastically reduces the 1-loop perturbative mixing between $\CO_{1-loop}(\textrm{taste P})$ and certain wrong-taste operators.  In particular, for the case of hypercubic fat (HYP) links, the $\CO(\alpha/4\pi)$ mixing with taste $S$ and taste $T$ operators is so small that we can consider it to be of higher-order in our power-counting \cite{LSMatch}.  This leaves the operators $\CO^{1I}_\chi$ and $\CO^{1T}_\chi$ each with only a single coefficient of $\CO(\alpha^2)$, thereby reducing the total number of fit parameters by two.\footnote{We presume that a similar simplification occurs for Asqtad links, although their perturbative matching coefficients have yet to be determined.}  

The biggest source of potential improvement, however, is in better perturbative matching.  A good first step would therefore be to match to \emph{all tastes of lattice operators} at 1-loop, since the requisite matching coefficients are known.  While it has been thought that ``wrong taste'' operator contributions to \bk would be less important than those from taste $P$ operators, our power-counting and operator enumeration show that this is not the case in general. Complete 1-loop matching would eliminate the 6 coefficients of $\CO(\alpha/4\pi)$ in the first column Table~\ref{tab:NLOCoeff}.  Moreover, 2-loop matching of all tastes, or better yet fully nonperturbative matching, would completely remove the first column in Table~\ref{tab:NLOCoeff} and eliminate the operators $\CO^{1I}_\chi$ and $\CO^{1T}_\chi$ altogether at this order, reducing the total number of fit parameters from 37 to 24.\footnote{Recall that, generically, use of the matching procedure outlined in Eq.~(\ref{eq:Zmua}) to a given order in perturbation theory requires knowledge of the anomalous dimension at one higher-order.  For example, 1-loop matching requires the 2-loop anomalous dimension.  Thus, while 2-loop matching would greatly reduce the number of coefficients in the NLO S$\chi$PT fit, the 3-loop anomalous dimension must also be calculated.}          

\section{Conclusions}
\label{sec:Conc}

We have calculated \bk to NLO in S$\chi$PT for quenched, PQ, and full QCD, the form of which is necessary for correct continuum and chiral extrapolation of staggered lattice data.  These results apply to both infinite and finite spatial volume.  The most general expression is that for a PQ theory with three sea quarks in which all valence and sea masses are different.   

This is the first calculation of a hadronic weak matrix element that is not a conserved current in staggered chiral perturbation theory.   It illustrates the generic drawback of staggered fermions -- large numbers of operators, both on the lattice and in the chiral effective theory, due to the additional taste degree-of-freedom.  For example, while \bk at 1-loop in continuum chiral perturbation theory is proportional to a single operator coefficient, the same quantity in staggered $\chi$PT comes from 13 operators, many of which have more than one distinguishable coefficient.  To address this concern we show explicitly how to determine 6 of these coefficients (at a single lattice spacing) from other simple matrix elements of the same lattice operator, thereby reducing the number of undetermined fit parameters.  Furthermore, we suggest two ways to reduce the large number of coefficients -- use of improved links, e.g. HYP or Asqtad, and of nonperturbative matching coefficients.  As usual, partial quenching helps to provide more handles in the fits, and will likely be essential here.

Use of our expression for \bk at NLO in S$\chi$PT, in combination with sufficient lattice data, should allow a precise determination of \bk with staggered quarks.  

\section*{Acknowledgments}
This work was supported by the U.S. Department of Energy under grant no. DE-FG02-96ER40956.

\appendix

\section{\bk in Continuum PQ$\chi$PT with 2+1 Sea Quarks}
\label{app:contPQ}

The continuum partially quenched result for $B_K$ for
$N_{sea}$ degenerate sea quarks has been given
by Golterman and Leung~\cite{gl}, and checked for $N_{sea}=2$ by
Becirevic and Villadoro~\cite{bv}. The result for ``$2+1$'' flavors
of sea quarks does not, however, exist in the literature.
It can be obtained by taking the appropriate limit of our staggered PQ expression, Eq.~(\ref{eq:BKPQ}), and here we present the result.

To go from Eq.~(\ref{eq:BKPQ}) to the continuum theory, we first set all coefficients associated with lattice errors, i.e. those multiplied by factors of $a^2$, $a_\alpha^2$, $\alpha/4\pi$, and $\alpha^2$, equal to zero.  This eliminates all operator coefficients but $\CC^K_\chi$.  It also removes the splittings among the sixteen tastes of PGBs;  we are therefore free to remove the taste subscripts.  Next we take the ratio of \bk to $B_0$ so that all factors of four due to traces in taste space drop out.  Finally we set $m_u = m_d \neq m_s$.  The next-to-leading order result for $B_K$ is then
\begin{equation}
\Bigg( \frac{B_K}{B_0} \Bigg)^\textrm{PQ, 2+1} = 1 + \frac{1}{48 \pi^2 f^2 m^2_{xy}}
\bigg[I_{conn}+I_{disc} + b m^4_{xy} + c (m^2_X-m^2_Y)^2 + d m^2_{xy}(2m^2_D+m^2_S)
\bigg]
\,,
\end{equation}
where $f\approx 130\;$MeV.  The coefficients
$b$, $c$ and $d$ are unknown dimensionless low-energy
constants proportional to the constants
$B$, $C$ and $D$ in Eq.~(\ref{eq:BKPQ}).
Note that the $d$ term is proportional to the sum of
the sea quark masses, and, to this order, effectively
renormalizes $B_0$ in a fit at a given sea quark mass.
Note also that $c$ is absent for degenerate valence quarks,
$m_x=m_y$.

The chiral logarithms are given by $I_{conn}$ and $I_{disc}$.
We have broken them into two contributions since the latter
vanishes when $m_x=m_y$ and is also the only place where
sea quark masses enter. The result from the quark connected
diagrams is
\begin{equation}
I_{conn} = 6 m^4_{xy} \tilde\ell(m^2_{xy}) - 3 \ell(m^2_X) (m^2_{xy} + m^2_X) - 3 \ell(m^2_Y) (m^2_{xy}+m^2_Y)
\,,
\end{equation}
while that for the diagrams involving a hairpin vertex is
\begin{eqnarray}
I_{disc} &=& I_X + I_Y + I_\eta\,,\\
I_X &=& \tilde\ell(m^2_X) \frac{(m^2_{xy}+m^2_X)(m^2_D-m^2_X)(m^2_S-m^2_X)}{(m^2_{\eta}-m^2_X)} \nonumber\\
 &-& \ell(m^2_X)\Bigg[\frac{(m^2_{xy}+m^2_X)(m^2_D-m^2_X)(m^2_S-m^2_X)}{(m^2_{\eta}-m^2_X)^2} + \frac{2(m^2_{xy}+m^2_X)(m^2_D-m^2_X)(m^2_S-m^2_X)}{(m^2_Y-m^2_X)(m^2_{\eta}-m^2_X)} \nonumber\\
&+& \frac{(m^2_D-m^2_X)(m^2_S-m^2_X)-(m^2_{xy}+m^2_X)(m^2_S-m^2_X)-(m^2_{xy}+m^2_X)(m^2_D-m^2_X)}{(m^2_{\eta}-m^2_X)}
 \Bigg]
,\\
I_Y &=& I_X(X\leftrightarrow Y)
\,,\\
I_\eta &=& \ell(m^2_{\eta}) \frac{(m^2_X-m^2_Y)^2 (m^2_{xy}+m^2_{\eta})(m^2_D-m^2_{\eta})(m^2_S-m^2_{\eta})}{
                           (m^2_X-m^2_{\eta})^2(m^2_Y-m^2_{\eta})^2}
\,.
\end{eqnarray}
The expression appears singular, but in fact is not --
the poles all cancel when $m^2_X,m^2_Y\to m^2_\eta$ or $m^2_X\to m^2_Y$.
Nevertheless, we have not found a simpler expression.
We have also checked that this expression becomes equal to the
two sea quark result (as given most compactly
in Becirevic and Villadoro~\cite{bv}) in the limit
that $m^2_S \gg m^2_D,m^2_X,m^2_Y,m^2_{xy}$ (so that $m^2_\eta=2 m^2_S/3$).

\section{Review of Staggered Chiral Perturbation Theory}
\label{app:SChPT}

Here we recall the leading-order effective staggered chiral Lagrangian, whose construction is discussed in much greater detail in Ref.~\cite{BA1}, and discuss some of its general consequences.  The next-to-leading order Lagrangian, determined in Ref.~\cite{Us}, turns out not to be necessary for \bk at NLO. 

\bigskip 

For $n$ staggered quark flavors, spontaneous breakdown of the approximate $SU(4n)$ chiral symmetry by the vacuum, 
\begin{equation}
	SU(4n)_L \times SU(4n)_R \rightarrow SU(4n)_V
\,,
\end{equation}
leads to $16n^2 - 1$ pseudo-Goldstone bosons (PGBs).  We collect them into an $SU(4n)$ matrix,    
\begin{equation}
\Sigma=\exp(i\Phi / f) \,,
\end{equation}
where $\Phi$ is a traceless $4n \times 4n$ matrix with $4 \times 4$ submatrices:
\begin{eqnarray}\label{eq:Phi}
	\Phi & = & \left( \begin{array}{cccc} U & \pi^+ & K^+ & \cdots \\*
	\pi^- & D & K^0 & \cdots \\* K^- & \bar{K^0} & S & \cdots \\*
	\vdots & \vdots & \vdots & \ddots \end{array} \right) \\
	U & = & \sum_{a=1}^{16} U_a T_a \,, \;\; \textrm{etc.}
\end{eqnarray}
and $f$ is normalized such that $f_\pi \approx 132\,$MeV.  We use Euclidean gamma matrices as the $SU(4)$ generators:
\begin{equation}\label{eq:T_a}
	T_a = \{ \xi_5, i\xi_{\mu 5}, i\xi_{\mu\nu}, \xi_{\mu},
	\xi_I\}\,,
\end{equation}
where $\xi_I$ is the $4 \times 4$ identity matrix.  The quark mass matrix is also of size $4n \times 4n$, but has trivial (singlet) taste structure:
\begin{eqnarray}
	\CM = \left( \begin{array}{cccc} m_u I & 0 &0 & \cdots \\* 0 &
	m_d I & 0 & \cdots \\* 0 & 0 & m_s I & \cdots\\* \vdots &
	\vdots & \vdots & \ddots \end{array} \right)\,.
\end{eqnarray}
Under chiral symmetry transformations, $\Sigma$ transforms as
\begin{eqnarray}
	&\Sigma \rightarrow  L \Sigma R^{\dagger}& \\
	&L \in SU(4n)_L , \;\;\; R \in SU(4n)_R .&
\end{eqnarray}
The standard S$\chi$PT power-counting scheme is\footnote{Note that standard S$\chi$PT does not include $\alpha$ in its power-counting because there are no external operators.}
\begin{equation}
	p^2 \approx m \approx a^2 \,,
\end{equation}
so the lowest order Lagrangian is of $\CO(p^2,m,a^2)$:
\begin{equation}
	\CL_\chi = \frac{f^2}{8} \Tr(\partial_{\mu}\Sigma
	\partial_{\mu}\Sigma^{\dagger}) - \frac{1}{4}\mu f^2 \Tr(\CM
	\Sigma + \CM \Sigma^{\dagger}) + \frac{2 m_0^2}{3}(U_I + D_I + S_I)^2 + a^2\CV\,,
\label{eq:ChLag}\end{equation}
where $\Tr$ is the full $4n \times 4n$ trace in both flavor and taste space, $\mu$ is a dimensionful constant of $\CO(\Lambda_{QCD})$, and $\CV$ is the taste symmetry breaking potential.

The taste breaking potential leads to important properties of staggered mesons.  For discussion of these features it is useful to separate $\CV$ into single and double trace operators:
\begin{eqnarray}
	\CV & = & \CU +\CU\,' 
\end{eqnarray}
\begin{eqnarray}
	-\CU & = & C_1
	\Tr(\xi^{(n)}_5\Sigma\xi^{(n)}_5\Sigma^{\dagger}) \nonumber
	\\* & & +C_3\frac{1}{2} \sum_{\nu}[ \Tr(\xi^{(n)}_{\nu}\Sigma
	\xi^{(n)}_{\nu}\Sigma) + h.c.] \nonumber \\* & &
	+C_4\frac{1}{2} \sum_{\nu}[ \Tr(\xi^{(n)}_{\nu 5}\Sigma
	\xi^{(n)}_{5\nu}\Sigma) + h.c.] \nonumber \\* & & +C_6\
	\sum_{\mu<\nu} \Tr(\xi^{(n)}_{\mu\nu}\Sigma
	\xi^{(n)}_{\nu\mu}\Sigma^{\dagger}) \label{eq:UPot} 
\end{eqnarray}
\begin{eqnarray}
	-\CU\,' & = &
	C_{2V}\frac{1}{4} \sum_{\nu}[ \Tr(\xi^{(n)}_{\nu}\Sigma)
	\Tr(\xi^{(n)}_{\nu}\Sigma) + h.c.] \nonumber \\*
	&&+C_{2A}\frac{1}{4} \sum_{\nu}[ \Tr(\xi^{(n)}_{\nu
	5}\Sigma)\Tr(\xi^{(n)}_{5\nu}\Sigma) + h.c.] \nonumber \\* & &
	+C_{5V}\frac{1}{2} \sum_{\nu}[ \Tr(\xi^{(n)}_{\nu}\Sigma)
	\Tr(\xi^{(n)}_{\nu}\Sigma^{\dagger})]\nonumber \\* & &
	+C_{5A}\frac{1}{2} \sum_{\nu}[ \Tr(\xi^{(n)}_{\nu5}\Sigma)
	\Tr(\xi^{(n)}_{5\nu}\Sigma^{\dagger}) ],
\end{eqnarray}
where $\xi_T^{(n)}$ is a $4n\times 4n$ matrix with an ordinary $4 \times 4$ taste matrix, $\xi_T$, repeated along the diagonal:
\begin{equation}
	\xi_T^{(n)} = \left( \begin{array}{cccc} \xi_T & 0 & 0 &
	\cdots\\* 0 & \xi_T & 0 & \cdots\\* 0 & 0 & \xi_T & \cdots \\*
	\vdots &\vdots & \vdots &\ddots\end{array} \right)\,.
\label{eq:xi_T}\end{equation}
The single trace operators in $\CU$ split the tree-level PGB masses into degenerate groups:
\begin{equation}
	\left(m_{\pi}^2\right)_{\rm LO}=\mu (m_i + m_j) + a^2 \Delta_F \,,
\label{eqn:LOmass}\end{equation}
where $\Delta_F$ is different for each of the five $SO(4)$-taste irreps $P$, $A$, $T$, $V$, and $I$.\footnote{We assume in this expression that the quark flavors $i$ and $j$ are different.}  The double trace operators in $\CU'$ generate hairpin diagrams (quark disconnected contractions) for flavor-neutral, taste vector and axial vector PGBs.  Both $\CU$ and $\CU'$ produce interaction vertices among even numbers of PGBs.  In addition, the $m_0^2$ term, present because only the \emph{overall} singlet can be integrated out of the theory, generates flavor-neutral, taste-singlet hairpins.  

\bigskip

It is easy to generalize the standard staggered chiral Lagrangian to quenched and partially quenched theories.  For a PQ theory with $N = n_{val} + n_{sea}$ quarks and $M = n_{val}$ ghosts, the chiral symmetry group is $SU(4N|4M)_L \times SU(4N|4M)_R$ and breaks to $SU(4N|4M)_V$.  The net result is that there are more PGBs, and all of the traces become supertraces ($\Tr \rightarrow \Str$) in the chiral Lagrangian. 

\section{Determination of Chiral \bk Operators}
\label{app:ops}

Here we determine all of the mesonic operators which contribute to $\CM_\textrm{lat}$, and consequently to $B_K$, at LO and NLO in our power-counting.  There are quite a few such operators because the continuum $SU(3)_{flavor}$ is enlarged to the graded, staggered symmetry $SU(4N_{val}+4N_{sea}|4N_{val})$ on the lattice.  Fortunately, the fact that $\CO_K$ is invariant under the lattice $U(1)_A$ symmetry protects the number of operators that contribute to \bk from becoming unwieldy.

We first enumerate the operators which contribute to \bk at tree-level.  These include the ``standard'' \bk chiral operator, and are relatively easy to determine, even in the staggered theory.  Next we enumerate the operators which contribute to \bk at 1-loop.  These can come either from operator-mixing on the lattice or from insertions (at the quark-level) of the staggered action with the \bk operator.  We then determine the analytic contributions to \bk at NLO, relying heavily on constraints from $U(1)_A$ and CPS symmetries.  Finally we discuss the parametric dependence of the various operator coefficients on the lattice spacing, $a$, and the strong coupling-constant, $\alpha$.  

\subsection{Operators that Contribute at Lowest Order}
\label{app:LO}

We must map the staggered quark-level operator, $\CO_K^{staggered}$, onto operators in the chiral effective theory.  This is done most simply by first splitting the operator into spin vector and spin axial-vector parts, mapping \emph{these} operators onto mesonic operators, and then taking appropriate linear combinations to get chiral operators which correspond to the staggered \bk operator.  

We first separate $\CO_K^{staggered}$ into two pieces\footnote{From now on we suppress color indices because both color contractions lead to the same chiral operators.}:
\begin{eqnarray}
	\CO_V & \equiv & [\bar{s1} (\gamma_\mu \otimes \xi_5\big) d1][\bar{s2} (\gamma_\mu \otimes \xi_5 ) d2] \\
	\CO_A & \equiv & [\bar{s1} (\gamma_\mu \gamma_5 \otimes \xi_5) d1][\bar{s2} (\gamma_\mu \gamma_5 \otimes \xi_5 ) d2] .
\end{eqnarray}
The \bk matrix element is proportional to their sum:
\begin{equation}
	\CM_K^{staggered} \propto < \bar{K^0_2}| (\CO_V + \CO_A) | K^0_1 > .
\end{equation}
As in Ref.~\cite{LS}, we label these four-fermion operators by the spin and taste in each bilinear, such that $\CO_V \propto [V \times P]$ and $\CO_A \propto [A \times P]$.  Note the convention that $A \equiv \gamma_{\mu 5} \otimes \gamma_{5 \mu}$, where the pair of spin (or taste) matrices are in separate bilinears, such that
\begin{equation}
	\CO_K^{staggered} \propto [V \times P] - [A \times P]
\end{equation}
in this condensed notation.

We now use spurion analysis as in Refs.~\cite{LS} and \cite{Us} to determine the resulting chiral operators.  The chiral structure of four-fermion operators with spins $V$ or $A$ is
\begin{equation}
\CO_F= \pm \sum_\mu \left(\bar {Q_R} (\gamma_\mu\otimes F_R) Q_R \pm 
\bar{Q_L} (\gamma_\mu\otimes F_L) Q_L \right)^2 \,,
\label{eq:OFform}
\end{equation}
where the upper and lower signs correspond to V and A, respectively, and $F_{L,R}$ are Hermitian taste matrices.  When we promote the taste matrices to spurion fields, $F_{L,R}$ must transform in the following way so that $\CO_F$ is invariant under chiral transformations:  
\begin{equation}
F_L\to L F_L L^\dagger, \;\;\;\;\; F_R \to R F_R R^\dagger \,.
\label{eq:VATrans}\end{equation}
We will set $F_L$ and $F_R$ to the appropriate matrices at the end.

We first note that chiral operators which result from two left-handed (or two right-handed) taste spurions have a \emph{relative minus sign} when they come from a spin $V$ versus a spin $A$ four-fermion operator, whereas those that result from one left-handed and one right-handed taste spurion do not.  Thus only operators built out of two spurions with the same handedness remain after taking the linear spin combination $(V-A)$ which corresponds to $\CO_K^{staggered}$.  There are no such operators without derivatives, and two such linearly independent ones with two derivatives.  These are shown in Table~\ref{tab:LOOps}.

\begin{table}\begin{tabular}{c}  \hline\hline

\multicolumn{1}{c}{\emph{Operator}}  \\[0.5mm] \hline

$\pm \Str(\Sigma \partial_\mu \Sigma^\dagger F_L \Sigma \partial_\mu\Sigma^\dagger F_L) \pm \Str(\Sigma^\dagger \partial_\mu \Sigma F_R \Sigma^\dagger \partial_\mu \Sigma F_R)$ \\[0.5mm]

$\pm \Str(\Sigma \partial_\mu \Sigma^\dagger F_L)\Str(\Sigma \partial_\mu \Sigma^\dagger F_L) \pm \Str(\Sigma^\dagger \partial_\mu \Sigma F_R)\Str(\Sigma^\dagger \partial_\mu \Sigma F_R)$ \\[0.5mm]

\hline\hline

\end{tabular}\caption{The two linearly-independent $\CO(p^2)$ operators corresponding to $[(V-A) \times F]$.  The $\pm$ signs correspond to spins $V$ and $A$, respectively.  Derivatives act only on the object immediately to the right.}\label{tab:LOOps}\end{table}

The expression in Eq.~(\ref{eq:OFform}) can be made to have the same flavor and taste structure as $\CO_K^{staggered}$ by setting the two spurions, which we now call $F_1$ and $F_2$, equal to sparse matrices with a single $\xi_5$ in either the $s1\bar{d1}$ or $s2\bar{d2}$ location.  With this replacement the first operator in Table~\ref{tab:LOOps} becomes
\begin{equation}
	\CO_{\chi}^{N} =  \sum_\mu \Big( \Str(\Sigma \partial_\mu \Sigma^\dagger F_1 \Sigma \partial_\mu\Sigma^\dagger F_2) + \Str(\Sigma^\dagger \partial_\mu \Sigma F_1 \Sigma^\dagger \partial_\mu \Sigma F_2) \Big) 
\label{eq:ON}\end{equation}
and the second becomes 
\begin{equation}
	\CO_{\chi}^{K} =  \sum_\mu \Big( \Str(\Sigma \partial_\mu \Sigma^\dagger F_1)\Str(\Sigma \partial_\mu \Sigma^\dagger F_2) + \Str(\Sigma^\dagger \partial_\mu \Sigma F_1)\Str(\Sigma^\dagger \partial_\mu \Sigma F_2) \Big).
\label{eq:OK}\end{equation}
Since $\CO_{\chi}^{K}$ has the same left-left current structure as the standard continuum full QCD operator \cite{BSW}, it should contribute to \bk at tree-level, and it does.  The matrix $F_1$ annihilates the $K^0_1$ (an $\bar{s1}d1$ meson) while $F_2$ produces the $\bar{K^0_2}$ (a $\bar{d2}s2$ meson). 

Because \bk in continuum $\chi$PT only depends on a single parameter \cite{BSW}, the coefficients of $\CO_\chi^K$ and $\CO_\chi^N$ cannot be independent.  We can determine their precise relationship using the fact that both operators in the chiral effective theory  arise from the same quark-level operator, $\CO_K^{staggered}$:
\begin{equation}
  \CO_K^{staggered} \xrightarrow[]{\textrm{S$\chi$PT}} \CC_{\chi}^{K}\CO_{\chi}^{K} + \CC_{\chi}^{N}\CO_{\chi}^{N},
\end{equation}
but they have nonvanishing tree-level matrix elements between different types of kaons.  In particular, while $\CO_{\chi}^{K}$ only has a tree-level matrix element between $\bar{K^0_1}$ and $K^0_2$ (by construction), $\CO_{\chi}^{N}$ only has one between ``mixed'' kaons, i.e. $s1\bar{d2}$ and $\bar{s2}d1$:
\begin{eqnarray}
	\langle\bar{K^0_1}_P | \CC_{\chi}^{K}\CO_{\chi}^{K} + \CC_{\chi}^{N}\CO_{\chi}^{N} | {K^0_2}_P\rangle & = & \frac{32 \CC_{\chi}^{K}}{f^2} \; m_{K}^2  \\ 
	\langle\bar{K^0_1}_P^{mix} | \CC_{\chi}^{K}\CO_{\chi}^{K} + \CC_{\chi}^{N}\CO_{\chi}^{N} | {K^0_2}_P^{mix}\rangle & = & \frac{8 m_K^2 C^N_\chi}{f^2}.
\end{eqnarray}
By comparing the same two matrix elements at the quark level we can extract the ratio $\CC_{\chi}^{K}:\CC_{\chi}^{N}$.  As written in Eq.~\ref{eq:OKStag}, $\CO_K^{staggered}$ can contract with $\bar{K^0_1}$ and $K^0_2$, but not the mixed kaons.  However, it can be Fierz-transformed into an operator with the appropriate mixed flavor structure, at the cost of introducing new tastes.  Using the Fierz-transformation rules in Appendix A of Ref.~\cite{SP}, we find that 
\begin{equation}
  [(V-A) \times P] \xrightarrow[]{\textrm{Fierz}} \frac{1}{4}[(V-A) \times (S+P+T-V-A)].
\end{equation}
Thus the ratio of the two matrix elements is 
\begin{equation}
  \frac{\langle\bar{K^0_1}_P | \CO_K^{staggered} | {K^0_2}_P\rangle}{\langle\bar{K^0_1}_P^{mix} | \CO_K^{staggered} | {K^0_2}_P^{mix}\rangle} = 4, 
\end{equation}
and $\CC_{\chi}^{K} = \CC_{\chi}^{N}$.  The Fierz transformation also reveals why the seemingly ``extra'' operator, $\CO_{\chi}^{N}$, is necessary in the chiral effective theory -- it accounts for the fact that $\CO_K^{staggered}$ has a nonvanishing tree-level matrix element between flavor-mixed kaons of \emph{all} tastes. 

\subsection{Operators that Contribute at NLO}
\label{app:NLO}

\bk receives NLO contributions from a host of additional operators because of lattice discretization effects.  In order to determine all such operators we must consider the three different ways in which they can arise at the quark-level:  through perturbative matching, mixing with higher-dimension operators, and insertions of operators from the staggered action.  

\subsubsection{1-Loop Matching Coefficients}
\label{app:1Loop}

Lattice operators can mix with each other as long as they are in the same representation of the symmetry group that maps a hypercube onto itself.  Consequently, lattice operators which correspond to continuum operators with a particular spin and taste \emph{mix} with other operators that correspond to a \emph{different continuum spin and taste}.  Specifically, although $\CO_K^{staggered}$ in the continuum is strictly a $[(V-A)\times P]$ four-fermion operator, once it is put on the lattice it mixes with $[(V+A)\times P]$ and other spin-taste structures.  Like all taste-breaking effects, this operator mixing is due to hard gluon exchange, so the amount of mixing can be calculated perturbatively in $\alpha$.  Such 1-loop ``matching coefficients" for staggered four-fermion operators have been calculated for both naive fermions and using hypercubic fat links \cite{LSMatch}.  Some of this operator mixing is accounted for and removed in present lattice calculations of \bk.  It is now standard to use 1-loop matching coefficients to remove the $\CO(\alpha/4\pi)$ mixing between the desired $[(V-A)\times P]$ lattice operator and the unwelcome $[(V+A)\times P]$ lattice operator.  This still leaves mixing at $\CO(\alpha^2)$ with $[(V+A)\times P]$ and at $\CO(\alpha/4\pi)$ with wrong taste operators.  We discuss them in turn.

Lattice operators with spin structure $(V+A)$ contribute to \bk, albeit suppressed by $\CO(\alpha^2)$.  Recall from the previous section that such operators arise from the combination of one left-handed spurion and one right-handed spurion, where $F_L$ and $F_R$ transform as in Eq.~(\ref{eq:VATrans}).  Here we need only consider operators without derivatives.  This is because any two-derivative operators are of $\CO(\alpha^2 p^2)$ and can only contribute to \bk at NLO through analytic terms, which we determine separately in Appendix~\ref{app:Anal}.  There is only one operator without derivatives, 
\begin{equation}	
	\CO_{\chi}^{1P} = \Str(F_1 \Sigma F_2 \Sigma^\dagger),  
\label{eq:O1P}\end{equation}
and it cannot contribute to \bk until 1-loop because of its flavor structure.  Note that the 1-loop contribution of $\CO_\chi^{1P}$ is of NLO in our power-counting because its coefficient is of $\CO(\alpha^2)$ and the loop momentum further suppresses it by $\CO(p^2)$.

Lattice calculations to date have not taken into account the $\CO(\alpha/4\pi)$ mixing between $[(V-A)\times P]$ and lattice operators of different tastes.  \bk therefore receives contributions from ``wrong taste'' lattice operators, but suppressed by $\CO(\alpha/4\pi)$.  Using the matching coefficients from Ref.~\cite{LSMatch} we find that the desired lattice operator mixes with the following eight four-fermion operators at 1-loop:
\begin{equation}
	[V \times S],\;\; [A \times S],\;\; [V \times T],\;\; [A \times T],\;\; [S \times V],\;\; [S \times A],\;\; [P \times V],\;\; [P \times A],\;\; [T \times V],\;\; [T \times A] ,
\label{eq:FFOps}\end{equation}
as well as taste off-diagonal operators, i.e. those in which the two bilinears have different tastes, and operators which violate rotational and $SO(4)$-taste symmetry.  We map all of these quark-level operators onto chiral operators using a straightforward generalization of the spurion analysis in Ref.~\cite{LS} to flavor off-diagonal operators.  Table~\ref{tab:TBOps} contains the results.  We do not show the chiral operators generated by $[T \times V,A]$ because they are simply a subset of those from $[S,P \times V,A]$.  Nor do we include taste off-diagonal operators, as they do not contribute to any PGB quantities of interest.  Furthermore, we neglect operators which break rotational and/or $SO(4)$-taste symmetry, as doing so would require at least four repeated Lorentz indices in the chiral operator.  This would have to come from the presence of four derivatives, four additional taste matrices, or two derivatives and two taste matrices, and therefore be of a higher order than we consider here.

\begin{table}\begin{tabular}{lcl}  \hline\hline

\multicolumn{1}{c}{\emph{Four-fermion Op.}} & \multicolumn{2}{c}{\emph{Chiral Op.}} \\[0.5mm] \hline

	$[V \times S] \textrm{ and } [A \times S]$ & $\longrightarrow$ & $\Str(\xi_{I} \Sigma \xi_{I} \Sigma^\dagger)$ \\[0.5mm]
	$[V \times T] \textrm{ and } [A \times T]$ & $\longrightarrow$ & $\Str(\xi_{\mu \nu} \Sigma \xi_{\nu \mu} \Sigma^\dagger)$ \\[0.5mm]
	$\textrm{}[S \times V] \textrm{ and } [P \times V]$ & $\longrightarrow$ & $\Str(\xi_\mu \Sigma)\Str(\xi_\mu \Sigma^\dagger)$ \\[0.5mm]
	$\textrm{}$ && $\Str(\xi_\mu \Sigma)\Str(\xi_\mu \Sigma) +  \Str(\xi_\mu \Sigma^\dagger)\Str(\xi_\mu \Sigma^\dagger)$ \\[0.5mm]
	$\textrm{}$ && $\Str(\xi_\mu \Sigma \xi_\mu \Sigma) + \Str(\xi_\mu \Sigma^\dagger \xi_\mu \Sigma^\dagger)$ \\[0.5mm]
	$\textrm{}[S \times A] \textrm{ and } [P \times A]$ & $\longrightarrow$ & $\Str(\xi_{\mu 5} \Sigma)\Str(\xi_{5 \mu} \Sigma^\dagger)$ \\[0.5mm]
	$\textrm{}$ && $\Str(\xi_{\mu 5} \Sigma)\Str(\xi_{5 \mu} \Sigma) +  \Str(\xi_{\mu 5} \Sigma^\dagger)\Str(\xi_{5 \mu} \Sigma^\dagger)$ \\[0.5mm]
	$\textrm{}$ && $\Str(\xi_{\mu 5} \Sigma \xi_{5 \mu} \Sigma) + \Str(\xi_{\mu 5} \Sigma^\dagger \xi_{5 \mu} \Sigma^\dagger)$ \\[0.5mm]

\hline\hline

\end{tabular}\caption{Mesonic operators corresponding to the four-fermion operators listed in Eq.~\ref{eq:FFOps}.  Repeated indices are summed with the constraint that $\mu \neq \nu$ in the taste tensor matrices.}\label{tab:TBOps}\end{table}

Because these operators arise from mixing with $\CO_K^{staggered}$, they have the \emph{same flavor structure} as the original continuum operator.  Thus $\xi_{\mu \nu} \rightarrow F_{1T}$ and $\xi_{\nu \mu} \rightarrow F_{2T}$, where $F_{1T}$ ($F_{2T}$) is a sparse matrix with a single $\xi_{\mu \nu}$ ($\xi_{\nu \mu}$) in the $s1\bar{d1}$ ($s2\bar{d2}$) location.  We define $F_{1I}$, $F_{1I}$, $F_{1V}$, $F_{2V}$, $F_{1A}$, and $F_{2A}$ analogously, such that the following eight operators contribute to $B_K$:\footnote{We note as an interesting aside that $\CO^{1I}_\chi$ does not appear in the staggered chiral Lagrangian because, if $\xi_I$ is flavor diagonal, it is just a trivial constant.}  
\begin{eqnarray}
	\CO_{\chi}^{1I} & = &  \sum_{\mu \neq \nu} \Str(F_{1I} \Sigma F_{2I} \Sigma^\dagger) \label{eq:O1I}\\
	\CO_{\chi}^{1T} & = &  \sum_{\mu \neq \nu} \Str(F_{1T} \Sigma F_{2T} \Sigma^\dagger) \label{eq:O1T}\\
	\CO_{\chi}^{1V} & = &  \sum_\mu \Big( \Str(F_{1V} \Sigma F_{2V} \Sigma) + \Str(F_{1V} \Sigma^\dagger F_{2V} \Sigma^\dagger) \Big) \\	
	\CO_{\chi}^{2V} & = &  \sum_\mu \Big( \Str(F_{1V} \Sigma)\Str(F_{2V} \Sigma) + \Str(F_{1V} \Sigma^\dagger)\Str(F_{2V} \Sigma^\dagger) \Big) \\
	\CO_{\chi}^{3V} & = &  \sum_\mu \Str(F_{1V} \Sigma)\Str(F_{2V} \Sigma^\dagger) \\
	\CO_{\chi}^{1A} & = &  \sum_\mu \Big( \Str(F_{1A} \Sigma F_{2A} \Sigma) + \Str(F_{1A} \Sigma^\dagger F_{2A} \Sigma^\dagger) \Big) \\	
	\CO_{\chi}^{2A} & = &  \sum_\mu \Big( \Str(F_{1A} \Sigma)\Str(F_{2A} \Sigma) + \Str(F_{1A} \Sigma^\dagger)\Str(F_{2A} \Sigma^\dagger) \Big) \\
	\CO_{\chi}^{3A} & = &  \sum_\mu \Str(F_{1A} \Sigma)\Str(F_{2A} \Sigma^\dagger)\,. \label{eq:O3A}
\label{eq:1Loop}\end{eqnarray}
Moreover, since Eq.~(\ref{eq:FFOps}) turns out to be a complete list of $U(1)_A$ and $SO(4)$-rotation invariant four-fermion operators, these are also the only wrong-taste operators that can contribute to \bk at $\CO(\alpha^2)$.  Thus all of the operators in Eq.~(\ref{eq:1Loop}) have two independent coefficients, one of $\CO(\alpha/4\pi)$ and the other of $\CO(\alpha^2)$.  Note that, as one would expect, such ``wrong taste" operators can only contribute to the \bk matrix element through loops.  Such contributions are of NLO in our power-counting because of the factor of $\alpha/4\pi$ or $\alpha^2$ in the coefficients and the further suppression of $\CO(p^2)$ by the loop-momentum.  We therefore do not need to consider wrong-taste operators with derivatives. 

\subsubsection{Dimension 8 Quark-level Operators}
\label{app:Dim8}

The four-fermion operator $\CO_K^{staggered}$ is dimension 6.  As discussed in the previous subsection, it mixes with other dimension 6 operators on the lattice, but such mixing is due to gluon exchange and suppressed by $\alpha/4\pi$.  Therefore the naively discretized version of the continuum operator, $\CO_K^{staggered,lat}$, generates the dominant contribution to the lattice $\bar{K^0_2}-K^0_1$ matrix element, $\CM_{lat}$.  However, many higher-dimension quark-level operators respect the \emph{same lattice symmetries} as $\CO_K^{staggered}$ and thus can also contribute to \bk on the lattice.  Dimension 7 and 8 quark-level operators are explicitly suppressed in the action relative to dimension 6 operators by factors of $a$ and $a^2$, respectively, so they can map onto chiral operators which generate 1-loop contributions to \bk at NLO.   In fact, because $a^2 \sim \alpha^2 \sim \alpha/4\pi$ in our power-counting, they are just as or more important than the dimension 6 operators which contribute through mixing.  Fortunately we need not go higher than dimension 8 at the quark-level because 1-loop contributions to \bk from dimension 9 operators are at least $\CO(a^3 p^2)$ and thus of higher than NLO.         

At the quark level, dimension 7 operators contain four fermions plus either a derivative or a factor of the quark mass matrix.  However, all dimension 7 quark level operators turn out to be \emph{taste off-diagonal} \cite{Sharpe:1993ng}, and therefore cannot contribute to any PGB processes of interest.  Thus we move on to dimension 8 quark-level operators, which contain four fermions plus two derivatives, two factors of the quark mass matrix, or one of each.  Dimension 8 operators with one derivative and one mass matrix must have a single nontrivial spin or taste matrix (either $V$ or $A$) in order to contract up the free derivative index.  Such operators are either spin off-diagonal or taste off-diagonal; both are irrelevant.  Dimension 8 operators with two mass matrices map onto chiral operators with at least two mass spurions.  Because they also carry a factor of $a^2$ from the action, such operators are of $\CO(a^2 m^2)$ or higher in our power-counting and can be neglected.  We therefore need only consider dimension 8 four-fermion operators with two derivatives that possess the same lattice symmetries as $\CO_K^{staggered}$.  Enumerating all such operators would be a tedious exercise in staggered group theory.  Fortunately, this exercise is unnecessary if we start with all \emph{chiral} operators which could contribute to \bk at NLO and work backwards.  

All chiral operators which arise from dimension 8 quark-level operators automatically come with a power of $a^2$.  Thus, in order to generate 1-loop contributions to \bk at NLO, they cannot contain any derivatives or be suppressed by additional powers of $a$ or $\alpha$.  All of the operators listed in Table~\ref{tab:TBOps} satisfy this condition.  In fact, it turns out that there are no additional chiral operators which are invariant under $U(1)_A$ and derivative-free. Therefore all of the chiral operators in Eqs.~(\ref{eq:O1P}) and (\ref{eq:O1I})-(\ref{eq:O3A}) should contribute to \bk both through 1-loop mixing, suppressed by $\CO(\alpha/4\pi)$, and through dimension 8 operators, suppressed by $\CO(a^2)$.   

\bigskip

Yet there is a subtlety -- not all $a^2$'s are created equal.  This is because, at the quark level, taste-changing four-fermion interactions require gluon exchange, and therefore have hidden powers of $\alpha$ which do not appear explicitly in the chiral effective Lagrangian, Eq.~(\ref{eq:ChLag}).  We must therefore consider the $\CO(``a^2")$ operators in more detail to allow correct extrapolation of lattice data.

In the case of matching the continuum version of $\CO_K^{staggered}$ onto dimension 8 lattice four-fermion operators, we want to find the dimension 8 operators that match onto with $[(V-A)\times P]$ \emph{without} gluon exchange, because they truly contribute at $\CO(a^2)$.  Any other dimension 8 operators contribute at $\CO(\alpha a^2)$ or higher, and thus are of at least NNLO in our power counting.  As discussed in both Refs.~\cite{DS} and \cite{PS}, lattice bilinears of different spins and tastes ``mix"\footnote{Recall that we are using the terminology ``mix" as shorthand for this process of matching lattice operators onto those in the continuum.} with each other at tree-level in perturbation theory, but to all orders in $a$, simply because of the fact that the spinor components of staggered fermions are spread out over a hypercube.\footnote{Note that these mixings are independent of both the choice of gauge links and the choice of how to construct gauge-invariant operators.}  Fortunately, at tree-level, one can consider the two bilinears in each four-fermion operator separately.  Thus, in order to determine which four-fermion operators mix at $\CO(a^2)$, one needs only find \emph{which bilinears mix at $\CO(1)$, $\CO(a)$, and $\CO(a^2)$}.  The product of two $\CO(a)$ bilinears, or of one $\CO(1)$ and one $\CO(a^2)$ bilinear, will be a four-fermion operator that mixes at $\CO(a^2)$.

We do so using Eqs.~(26) and (27) in Ref.~\cite{PS}.  We find that the dimension 3 bilinear $[\gamma_\mu(1 + \gamma_5) \otimes \xi_5]$ mixes with taste $V$ dimension 4 bilinears at $\CO(a)$.  It also mixes with dimension 3 and 5 taste $P$ bilinears at $\CO(1)$ and $\CO(a^2)$, respectively.  Because spin off-diagonal four-fermion operators cannot contribute to PGB processes, we are restricted to multiplication of each of these bilinears by itself.  For example, we can multiply the bilinear $[1 \times \xi_\mu]$ times itself to generate the four-fermion operator  
\begin{equation}
	\sum_\mu \sum_\nu \; \bar{Q}D_\mu(I \otimes \xi_\mu)Q \; \bar{Q}D_\nu(I \otimes \xi_\nu)Q\,.
\label{eq:Dim8}\end{equation}
The above operator is simply a generalization of the dimension 6 four-fermion operator $[S \times V]$ to dimension 8.  The derivatives give it the appropriate dimension, and allow it to have both $SO(4)$ rotational symmetry breaking ($\mu = \nu$) and conserving ($\mu \neq \nu$) pieces.     

It is clear now that the lattice version of $\CO_K^{staggered}$ really only mixes at $\CO(a^2)$ with dimension 8 quark-level operators that have spin-taste structures $[S \times V]$, $[P \times V]$, $[T \times V]$, $[(V-A) \times P]$, and $[(V+A) \times P]$ along with two derivatives.  These lead to the four chiral operators $\CO^{1V}_\chi$, $\CO^{2V}_\chi$, $\CO^{3V}_\chi$, and $\CO^{1P}_\chi$.\footnote{They also lead to higher-order chiral operators which break $SO(4)$ taste symmetry, corresponding to the piece of Eq.~(\ref{eq:Dim8}) with $\mu = \nu$.}  We therefore conclude that their coefficients are of $\CO(a^2)$, while the rest of the operators enumerated this section can be neglected.  This is summarized in the second column of Table~\ref{tab:NLOCoeff}.    

\subsubsection{Insertions of the Staggered Action}
\label{app:Act}

\begin{figure}
	\epsfysize=0.73in \epsffile{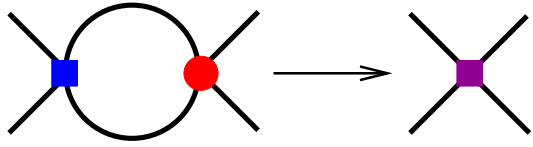}
	\caption{Schematic of \bk vertex modification due to the insertion of an operator from the staggered action.  In the first diagram the square represents the original \bk vertex and the circle represents the inserted strong four fermion vertex.  In the second diagram the square represents the resulting local vertex in the effective theory.}\label{fig:ActIns}
\end{figure}

The lattice operator mixing discussed in the past two subsections is not particularly intuitive because it is a stems from perturbation theory and complicated lattice symmetries.  In contrast, the fact that operators from the staggered action modify the quark-level \bk operator, $\CO_K^{staggered}$, can be shown through a simple diagrammatic argument.  Consider the four-fermion $\CO_K^{staggered}$ and another four-fermion operator from the staggered action.  Two of the quarks in $\CO_K^{staggered}$ can contract with two of the quarks in the other operator to form a 1-loop diagram with four external quarks, as shown in Fig.~\ref{fig:ActIns}.  This produces a local operator in the chiral effective theory because the loop propagates over $\CO(1/\Lambda_{QCD})$, rather than over $\CO(1/m_\pi)$.  It corrects $\CO_K^{staggered}$ at $\CO(a_\alpha^2)$ because all of the dimension 6 four-fermion operators in the staggered action are taste-breaking.  Such $\CO(a_\alpha^2)$ corrections turn out to be numerically enhanced, however, so they are still of NLO in our power-counting.  We must therefore determine all chiral operators produced by combining $\CO_K^{staggered}$ with each of the dimension 6 four-fermion operators in the staggered action.

Ref.~\cite{Us} developed a method for combining four-fermion operators at the chiral level.  Essentially, one forms chirally invariant operators which contain two taste spurions from the first four-fermion operator and two from the second.  $\CO_K^{staggered}$ has spin-taste structure $[(V-A) \times P]$, so the two taste spurions from $\CO_K^{staggered}$ must transform as in Eq.~(\ref{eq:VATrans}).  Moreover, we learned in Sec.~\ref{app:LO} that only operators with two left-handed (or two right-handed) taste spurions survive after taking the linear spin combination $(V-A)$.  In contrast, the four-fermion operator may have any spin -- $V, A, S, P,$ or $T$ -- combined with an appropriate taste to maintain overall $U(1)_A$-invariance.  Furthermore, the two taste spurions from the inserted four-fermion operator must be \emph{flavor diagonal} because the action does not change quark flavor.    

Operators with spins $V$ or $A$ can have tastes $S, P,$ or $T$, and their taste spurions transform as Eq.~(\ref{eq:VATrans}).  Generically, there are three possible ways of combining these spurions with those from $\CO_K^{staggered}$:      
\begin{eqnarray}
	&&(F_L F_L)(F_L' F_R') + p.c. \nonumber \\ 
	&&(F_L F_L)(F_L' F_L') + p.c. \nonumber \\ 
	&&(F_L F_L)(F_R'F_R') + p.c. , 
\end{eqnarray}
where ``p.c.'' indicates parity-conjugate.  Here the unprimed spurions are from $\CO_K^{staggered}$ and the primed are from the four-fermion operator in the staggered action.  Because the two $F_L$'s will ultimately have different off-diagonal flavor structures, they must be separated by at least one $\Sigma$-field on either side within a supertrace.  We use this fact to eliminate a few operators.  Consequently, these spurion combinations only lead to two nontrivial generic chiral structures, shown at the top Table~\ref{tab:InsOps}.   To produce operators which contribute to \bk we let $F_L \rightarrow F_1 \& F_2$ and $F_R' \rightarrow \xi_I, \xi_5,$ or $\xi_{\mu \nu}$. It is easy to see that the chiral operators are trivial when the taste spurions from the quark-level operator in the action are just the identity.  Thus we are left with four new operators which contribute to \bk with coefficients of $\CO(a_\alpha^2)$:
\begin{eqnarray}	
	\CO_{\chi}^{2P} &=&  \Str(\xi_5  \Sigma F_1 \Sigma^\dagger \xi_5  \Sigma F_2 \Sigma^\dagger) + p.c. \label{eq:O2P} \\
	\CO_{\chi}^{3P} &=&  \Str(\xi_5 \Sigma F_1 \Sigma^\dagger)\Str(\xi_5 \Sigma F_2 \Sigma^\dagger) + p.c. \\
	\CO_{\chi}^{2T} &=&  \sum_{\mu \neq \nu} \Str(\xi_{\mu \nu}  \Sigma F_1 \Sigma^\dagger \xi_{\nu \mu}  \Sigma F_2 \Sigma^\dagger) + p.c. \\
	\CO_{\chi}^{3T} &=&  \sum_{\mu \neq \nu} \Str(\xi_{\mu \nu} \Sigma F_1 \Sigma^\dagger)\Str(\xi_{\nu \mu} \Sigma F_2 \Sigma^\dagger) + p.c.\,. \label{eq:O3T}
\end{eqnarray}       

\begin{table}\begin{tabular}{lcl}  \hline\hline

\multicolumn{1}{c}{\emph{Generic Structure}} & \multicolumn{2}{c}{\emph{Specific Operator}} \\[0.5mm] \hline

	$\Str(F_L \Sigma F_R' \Sigma^\dagger F_L \Sigma F_R' \Sigma^\dagger) + p.c.$ & $\longrightarrow$ & $\Str( F_1  F_2 ) + p.c.$ \\[0.5mm]
	$\textrm{}$ && $\Str(\xi_5  \Sigma F_1 \Sigma^\dagger \xi_5  \Sigma F_2 \Sigma^\dagger) + p.c.$ \\[0.5mm]
	$\textrm{}$ && $\Str(\xi_{\mu \nu} \Sigma F_1 \Sigma^\dagger \xi_{\nu \mu} \Sigma F_2 \Sigma^\dagger) + p.c.$ \\[0.5mm]
	$\Str(F_L \Sigma F_R' \Sigma^\dagger)\Str(F_L \Sigma F_R'\Sigma^\dagger) + p.c.$ & $\longrightarrow$ & $\Str( F_1 )\Str(F_2 ) + p.c.$ \\[0.5mm]
	$\textrm{}$ && $\Str(\xi_5 \Sigma F_1 \Sigma^\dagger)\Str(\xi_5 \Sigma F_2 \Sigma^\dagger) + p.c.$ \\[0.5mm]
	$\textrm{}$ && $\Str(\xi_{\mu \nu} \Sigma F_1 \Sigma^\dagger)\Str(\xi_5 \Sigma \xi_{\nu \mu} \Sigma^\dagger) + p.c.$ \\[0.5mm]

	\hline

	$\Str(F_L \tilde{F_L} \Sigma^\dagger F_L \Sigma \tilde{F_R}) + p.c.$ & $\longrightarrow$ & $\Str(F_1 \xi_{\mu} \Sigma^\dagger F_2 \Sigma \xi_{\mu}) + p.c. $ \\[0.5mm]
	$\textrm{}$ && $\Str(F_1 \xi_{\mu 5} \Sigma^\dagger F_2 \Sigma \xi_{5 \mu}) + p.c.$ \\[0.5mm]
	$\Str(F_L \tilde{F_L} \Sigma^\dagger) \Str(F_L \Sigma \tilde{F_R}) + p.c.$ & $\longrightarrow$ & $\Str(F_1 \xi_\mu \Sigma^\dagger) \Str(F_2 \Sigma \xi_\mu) + p.c.$ \\[0.5mm]
	$\textrm{}$ && $\Str(F_1 \xi_{\mu 5} \Sigma^\dagger) \Str(F_2 \Sigma \xi_{5 \mu}) + p.c.$ \\[0.5mm]
	$\Str(F_L \tilde{F_L} \Sigma^\dagger F_L \tilde{F_L} \Sigma^\dagger) + p.c.$ & $\longrightarrow$ & $\Str(F_1 \xi_\mu \Sigma^\dagger F_2 \xi_\mu \Sigma^\dagger) + p.c.$ \\[0.5mm]
	$\textrm{}$ && $\Str(F_1 \xi_{\mu 5} \Sigma^\dagger F_2 \xi_{5 \mu} \Sigma^\dagger) + p.c.$ \\[0.5mm]
	$\Str(F_L \tilde{F_L} \Sigma^\dagger) \Str(F_L \tilde{F_L} \Sigma^\dagger) + p.c.$ & $\longrightarrow$ & $\Str(F_1 \xi_\mu \Sigma^\dagger) \Str(F_2 \xi_\mu \Sigma^\dagger) + p.c.$ \\[0.5mm]
	$\textrm{}$ && $\Str(F_1 \xi_{\mu 5} \Sigma^\dagger) \Str(F_2 \xi_{5 \mu} \Sigma^\dagger) + p.c.$ \\[0.5mm]

\hline\hline

\end{tabular}\caption{Mesonic operators corresponding to insertions of four-fermion operators from the staggered action.  $F_1$ and $F_2$ are flavor off-diagonal and come from the \bk quark-level operator.  The remaining taste matrices are flavor-diagonal.  Repeated indices are summed with the constraint that $\mu \neq \nu$ in the taste tensor matrices.  The notation ``p.c.'' indicates the parity-conjugate of the previous operator.  The difference between upper and lower panels is described in the text. }\label{tab:InsOps}\end{table}

Operators with spins $S$ or $P$ can have tastes $V$ or $A$.\footnote{We choose to neglect spin $T$ in this discussion because it ultimately leads to a subset of operators already generated by spins $S$ and $P$.}  Their chiral structure is as follows:
\begin{equation}
\CO'_F= \left(\bar {Q_L} (1 \otimes \tilde{F}_L) Q_R \pm 
\bar{Q_R} (1 \otimes \tilde{F}_R) Q_L \right)^2 \,,
\label{eq:OFPform}
\end{equation}
where and the upper and lower signs correspond to spin S and P, respectively, so their taste spurions must transform as
\begin{equation}
\tilde{F}_L\to L \tilde{F}_L R^\dagger, 
\;\;\;\;\; \tilde{F}_R \to R \tilde{F}_R L^\dagger \,.
\end{equation}
As before, there are three possible ways of combining these spurions with those from $\CO_K^{staggered}$:  
\begin{eqnarray}
	&&(F_L F_L)(\tilde{F_L} \tilde{F_R}) + p.c. \nonumber \\ 
	&&(F_L F_L)(\tilde{F_L} \tilde{F_L}) + p.c. \nonumber \\ 
	&&(F_L F_L)(\tilde{F_R} \tilde{F_R}) + p.c.\,.
\end{eqnarray}
These spurion combinations generate the four nontrivial chiral structures shown in the lower panel of Table~\ref{tab:InsOps}.  In this case, to produce operators which contribute to \bk we let $F_L \rightarrow F_1 \& F_2$ and $\tilde{F_L}, \tilde{F_R} \rightarrow \xi_\mu$ or $\xi_{\mu 5}$.  Because two of the taste matrices are diagonal, the following (and similar) relations allow further simplification of the resulting operators:
\begin{eqnarray}
	\xi_\mu F_1 \xi_\mu &=& -F_1 \nonumber \\
	\xi_\mu F_1 &=& F_{1A} \nonumber \\
	\xi_{\mu 5} F_1 &=& F_{1V} .
\end{eqnarray}	
Such simplification reveals that these operators are none other than other than $\CO_\chi^{1P}$ and $\CO_\chi^{1V}$ -- $\CO_\chi^{3A}$.  However, here they have arisen through insertions of the staggered action rather than lattice operator mixing.

In principle, we must also consider insertions of the dimension 4 quark mass term, which can modify $O_K^{staggered}$ in a similar manner to the four-fermion operators already considered.  The resulting chiral operators would contain the two taste spurions from $\CO_K^{staggered}$ and a mass spurion, and thus be of $\CO(m)$ in our power-counting.  However, it turns out that there are no such operators for quite a simple reason.  Recall that, since $F_1$ and $F_2$ are flavor off-diagonal, they must either be separated by a $\Sigma$-field on both sides or be in different supertraces, each with a $\Sigma$.  A single mass spurion is insufficient for meeting this criterion.  Therefore operators with mass spurions only contribute to \bk at NLO through analytic terms, which we discuss next.   

\subsection{Analytic NLO Contributions}
\label{app:Anal}

We finally consider analytic NLO contributions to the $B_K$ matrix element, $\CM_K$.  Many operators generate contributions with the same quark mass dependence; it is unnecessary to separate them in fits to lattice data.  Thus rather than enumerating all possible operators, we can determine all linearly-independent functions of the quark masses, allowing symmetries to do much of the work.

Given our power-counting scheme, generically six types of operators can contribute to $\CM_K$:  
\begin{eqnarray}
	\CO(p^4), \;\; \CO(``a^2" p^2), \;\; \CO(``a^4"), \;\; \CO(m^2), \;\; \CO(p^2 m), \;\; \CO(``a^2" m), 
\end{eqnarray}   
where quotation marks indicate that $a^2$ can be interchanged with either $\alpha^2$, $\alpha/4\pi$, or $a_\alpha^2$.  We can immediately rule out contributions from $\CO(p^4)$ operators because there are four derivatives but only two fields upon which they can act.  We can also rule out NLO analytic contributions from $\CO(``a^4")$ operators.  Without derivatives or mass matrices the only possible matrix element from such operators is $\propto ``a^4"$.  However, this contribution does not vanish in the chiral limit, and therefore violates the $U(1)_A$ symmetry possessed by $\CO_K^{staggered}$, so it cannot occur.  Effectively, $U(1)_A$ forces all tree-level matrix elements between taste-\emph{$\xi_5$} kaons to vanish in the chiral limit.\footnote{Of course, this is nothing more than a restatement of the fact that taste-$\xi_5$ mesons are lattice Goldstone bosons, but we have nevertheless checked this result explicitly.}   

Determination of the NLO analytic contributions to $\CM_K$  from $\CO(``a^2" p^2)$ operators is also straightforward.  The derivatives \emph{must} act on the two external kaons, bringing down a factor of $p_K^2$ which becomes $m_K^2$ when the kaons are on-shell.  Therefore the first NLO analytic term in the expression for $\CM_K$ is
\begin{equation}
	A m_K^2,
\end{equation}
where $A$ is an undetermined coefficient that is, in principle, suppressed by either $a^2$, $\alpha^2$, $\alpha/4\pi$, or $a_\alpha^2$.  For fitting it is important to know whether or not $A$ does, in fact, depend upon all four expansion parameters.  Operators of $\CO(a^2 p^2)$ come from mixing with dimension 8 operators.  Recall from Appendix~\ref{app:Dim8} that $[(V-A)\times P]$ mixes with taste $P$ and taste $V$ operators at $\CO(a^2)$.  The operator $\CO^K_\chi$ itself comes from a taste $P$ quark-level operator, so its coefficient, $\CC^K_\chi$, receives a correction of $\CO(a^2)$.  One can also easily map the quark-level operator $[(V+A)\times P]$ onto a two-derivative chiral operator that contributes to $\CM_K$ at tree-level:
\begin{equation}
  \Str(\Sigma  \partial_\mu \Sigma^\dagger F_1) \Str(\Sigma^\dagger \partial_\mu \Sigma F_2).
\label{eq:AmK}\end{equation}
Although we lump all NLO analytic contributions to $\CM_K$ of the form $a^2 m_K^2$ together, we note that it \emph{would} be important to separate the correction to $\CC^K_\chi$ from the rest if one were to do a fit including some NNLO terms.  Operators of $\CO(\alpha^2 p^2)$ come from perturbative matching with all tastes of four-fermion operators on the lattice, which we discussed in Appendix~\ref{app:1Loop}.  As we just showed, mixing with taste $P$ four-fermion operators leads both to corrections to $\CC^K_\chi$ and to a new operator.  This time they produce contributions to $\CM_K$ of the form $\alpha^2 m_K^2$.  In contrast, operators of $\CO(p^2 \alpha/4\pi)$ only come from perturbative matching with \emph{wrong-taste} dimension 6 four-fermion operators.  Because wrong-taste operators cannot contribute to $\CM_K$ at tree-level, however, there is no analytic contribution of the form $m_K^2 \alpha/4\pi$.  Finally, operators of $\CO(p^2 \alpha^2)$ come from insertions with operators in the action.  The insertion of $\CO_K^{staggered}$ with the taste-singlet dimension 6 four-fermion operator maps onto $\CO_\chi^K$, but now at higher-order.  Thus there is an NLO analytic correction to $\CC^K_\chi$, and therefore a contribution to $\CM_K$, of the form $a_\alpha^2 m_K^2$.  We summarize the parametric dependence of the coefficient $A$ in Table~\ref{tab:AnalCoeff}.        

Determining the analytic NLO contributions to $\CM_K$ from the last three types of operators requires a bit more work.  To do this, we utilize a symmetry of $\CO_K^{staggered}$ which has not been explicitly needed until now:  \emph{CPS symmetry} \cite{BDS}.  Recall that, because $\CO_K^{staggered}$ is a weak operator, it is \emph{not} parity-invariant, as is the strong chiral Lagrangian.  However, it \emph{is} invariant under CPS symmetry, and this can be used to restrict the chiral operators to which is corresponds. Standard CPS symmetry is CP plus $d \leftrightarrow s$ and $m_d \leftrightarrow m_s$.  In our case, because there are two sets of valence quarks, we must impose a modified CPS symmetry with both $d1 \leftrightarrow s1$ and $d2 \leftrightarrow s2$ so as to make the quark-level operator invariant.  At the chiral level, this corresponds to the following transformations:
\begin{eqnarray*}
	\mbox{C}: && \Sigma \leftrightarrow \Sigma^T \\
	\mbox{P}: && \Sigma \leftrightarrow \Sigma^\dagger \\
	\mbox{``S''}: && d1 \leftrightarrow s1,\;\; d2 \leftrightarrow s2 , 
\end{eqnarray*}   
where the last line denotes an interchange of indices in both $\Sigma$ and $M$.  One can easily show that all of the operators enumerated in the previous subsections are already invariant under this series of transformations.  However, CPS symmetry will prove important for restricting the analytic contributions to $\CM_K$.

First consider analytic contributions to $\CM_K$ from $\CO(m^2)$ operators, which necessarily contain the two spurions from $\CO_K^{staggered}$ plus two mass spurions.  Because of the \emph{flavor} structure of $F_1$ and $F_2$, such operators will only produce nonvanishing tree-level matrix elements if one mass spurion is within each supertrace, e.g.:
\begin{eqnarray}
	&&\Str(F_1 \Sigma M^\dagger)\Str(F_2 \Sigma M^\dagger) + \Str(F_1 M \Sigma^\dagger)\Str(F_2 M \Sigma^\dagger)\\
	&&\Str(F_1 \Sigma M^\dagger)\Str(F_2 M \Sigma^\dagger) + \Str(F_1 M \Sigma^\dagger)\Str(F_2 \Sigma M^\dagger)\,.
\end{eqnarray}
Note that the second term in each operator is the CPS conjugate of the first term.  When contracted with an external $\bar{K^0_2}$ and a $K^0_1$, the first operator generates an analytic term proportional to $m_d^2 + m_s^2$, while the second generates one proportional to $m_d m_s$.  We choose to combine these terms such that the new analytic contributions to $\CM_K$ at NLO are of the form
\begin{eqnarray}
	B m_K^4, && C (m_d - m_s)^2 .
\end{eqnarray}
Note that the second term vanishes quadratically as $(m_d - m_s) \rightarrow 0$.  

Next consider $\CO(p^2 m)$ operators which consist of the two spurions from $\CO_K^{staggered}$, two Lie derivatives, and a mass spurion.  To produce a nonvanishing matrix element, each Lie derivative must be in a separate supertrace with either $F_1$ or $F_2$, e.g.:
\begin{eqnarray}
	&&\Str(F_1 \Sigma \partial_\mu \Sigma^\dagger)\Str(F_2 \Sigma \partial_\mu \Sigma^\dagger)\Str(\Sigma M^\dagger + M \Sigma^\dagger)\,, \\
	&&\Str(F_1 \Sigma \partial_\mu \Sigma^\dagger)\Str(F_2\Sigma \partial_\mu \Sigma^\dagger \Sigma M^\dagger) + \Str(F_1 \Sigma \partial_\mu \Sigma^\dagger)\Str(F_2 M \Sigma^\dagger \Sigma \partial_\mu \Sigma^\dagger)\,.
\end{eqnarray}
Once again, these operators are CPS-invariant.  The first operator generates a term proportional to $m_K^2 \Str(M) = m_K^2 \Tr(M_{sea})$, while the second generates one proportional to $m_K^2(m_d + m_s) \propto m_K^4$.  Thus there is only one new analytic contribution to $\CM_K$ at NLO from such operators:
\begin{eqnarray}
	D m_K^2 \Tr(M_{sea}).
\end{eqnarray}

Lastly we consider $\CO(``a^2" m)$ operators.  These contain two spurions from $\CO_K^{staggered}$, possibly two flavor-diagonal taste spurions, and a mass spurion.  Thus their tree-level matrix elements are proportional to one power of the quark mass, for which there are three possibilities: $(m_d + m_s)$, $(m_d - m_s)$, and $\Tr(M_{sea})$.  However, $``a^2"(m_d - m_S)$ is not CPS-invariant, and $``a^2"(m_d + m_s)$ is included in the parameter $A$, so we need only consider analytic terms of the form $``a^2" \Tr(M_{sea})$.  The factor $\Tr(M_{sea})$ always arises from the operator $\Str(\Sigma M^\dagger + M \Sigma^\dagger)$ when $\Sigma = 1$, so the only operators that can produce tree-level contributions of the form $``a^2" \Tr(M_{sea})$ are simply the previously enumerated $\CO(a^2)$, $\CO(\alpha^2)$, $\CO(\alpha/4\pi)$, or $\CO(a_\alpha^2)$ operators multiplied by $\Str(\Sigma M^\dagger + M \Sigma^\dagger)$.  However, $\CO(``a^2")$ operators cannot generate tree-level contributions to $\CM_K$ because of the $U(1)_A$ symmetry, so neither do these particular $\CO(``a^2" m)$ ones.       

\subsection{Parametric Dependence of Operator Coefficients}
\label{app:Coeff}

Many of the chiral operators enumerated in the previous sections come from multiple quark-level operators.  Thus they can have \emph{multiple coefficients}, each multiplied by a different expansion parameter.  While this does not affect the form of \bk at NLO in S$\chi$PT, it certainly affects extrapolation of lattice data using multiple lattice spacings.  Thus, for ease of use, we summarize the dependence of the NLO operator coefficients on the various expansion parameters in Table~\ref{tab:NLOCoeff}.  All operators which come from 1-loop operator mixing have coefficients of $\CO(\alpha^2)$, $\CO(\alpha/4\pi)$, or both, as described in Section~\ref{app:1Loop}.  Operators which arise through mixing with dimension 8 four-fermion operators come in at $\CO(a^2)$.  Finally, operators that arise from insertions of four-fermion operators are of $\CO(\alpha^2 a^2)$.  Thus the various \bk operators have anywhere between 1 and 4 independent undetermined coefficients.  For example, $\CO_\chi^{1V}$ has four coefficients, multiplied by $\alpha/4\pi$, $\alpha^2$, $a^2$, and $a_\alpha^2$, respectively, whereas $\CO_\chi^{2P}$ has a single coefficient multiplied by $a_\alpha^2$.    

\begin{table}\begin{tabular}{lccc}  \hline\hline

\multicolumn{1}{c}{\emph{\bk Operator}} & \multicolumn{3}{c}{\emph{Source -- Coefficient Dependence}} \\[0.5mm] 

&\mbox{    }1-loop Mixing?\mbox{  }&\mbox{  }Dimension 8 Op.?\mbox{  }&\mbox{  }Action Insertion?\mbox{  } \\[0.5mm] \hline

$\CO_\chi^{1P}$& Yes -- $\alpha^2$ & Yes -- $a^2$ & Yes -- $a_\alpha^2$ \\[0.5mm] 
$\CO_\chi^{1T}$& Yes -- $\alpha/4\pi$\,, $\alpha^2$ & No & No \\[0.5mm] 
$\CO_\chi^{1V}$& Yes -- $\alpha/4\pi$\,, $\alpha^2$ & Yes -- $a^2$ & Yes -- $a_\alpha^2$ \\[0.5mm] 
$\CO_\chi^{2V}$& Yes -- $\alpha/4\pi$\,, $\alpha^2$ & Yes -- $a^2$ & Yes -- $a_\alpha^2$ \\[0.5mm] 
$\CO_\chi^{3V}$& Yes -- $\alpha/4\pi$\,, $\alpha^2$ & Yes -- $a^2$ & Yes -- $a_\alpha^2$ \\[0.5mm] 
$\CO_\chi^{1A}$& Yes -- $\alpha/4\pi$\,, $\alpha^2$ & No & Yes -- $a_\alpha^2$ \\[0.5mm] 
$\CO_\chi^{2A}$& Yes -- $\alpha/4\pi$\,, $\alpha^2$ & No & Yes -- $a_\alpha^2$ \\[0.5mm] 
$\CO_\chi^{3A}$& Yes -- $\alpha/4\pi$\,, $\alpha^2$ & No & Yes -- $a_\alpha^2$ \\[0.5mm] 
$\CO_\chi^{1I}$& Yes -- $\alpha/4\pi$\,, $\alpha^2$ & No & No  \\[0.5mm] 
$\CO_\chi^{2P}$& No & No & Yes -- $a_\alpha^2$ \\[0.5mm] 
$\CO_\chi^{3P}$& No & No & Yes -- $a_\alpha^2$ \\[0.5mm] 
$\CO_\chi^{2T}$& No & No & Yes -- $a_\alpha^2$ \\[0.5mm] 
$\CO_\chi^{3T}$& No & No & Yes -- $a_\alpha^2$ \\[0.5mm]  

\hline\hline

\end{tabular}\caption{Parametric dependence of NLO \bk operator coefficients.  Each chiral operator can come from more than one quark-level operator, and therefore enter the NLO expression for \bk multiplied by more than one expansion parameter.}\label{tab:NLOCoeff}\end{table}      

\bigskip 

Finally, for completeness, we include the dependence of the various analytic terms on $a$ and $\alpha$ in Table~\ref{tab:AnalCoeff}.  

\begin{table}[h]\begin{tabular}{lc}  \hline\hline

\multicolumn{1}{c}{\emph{\bk Analytic Term}} & \multicolumn{1}{c}{$\;\;\;\;\;$\emph{Coefficient Dependence}} \\[0.5mm] \hline

$A m_K^2$ & $a^2, \alpha^2, a_\alpha^2$  \\[0.5mm] 
$B m_K^4$ & $1$  \\[0.5mm] 
$C (m_d - m_s)^2$ & $1$  \\[0.5mm] 
$D\, m_K^2 \Tr(M_{sea})$ & $1$  \\[0.5mm]  

\hline\hline

\end{tabular}\caption{Parametric dependence of NLO analytic contributions to \bk. An entry of ``1'' indicates that the coefficient is a constant at this order.}\label{tab:AnalCoeff}\end{table}

\end{document}